\documentclass[fleqn,usenatbib]{mnras}

\usepackage[T1]{fontenc}

\DeclareRobustCommand{\VAN}[3]{#2}
\let\VANthebibliography\thebibliography
\def\thebibliography{\DeclareRobustCommand{\VAN}[3]{##3}\VANthebibliography}

\usepackage{graphicx}	
\usepackage{amsmath}	
\usepackage{amssymb}	

\graphicspath{{./}{figures/}}
\usepackage{color,subfigure,lineno} 

\usepackage{eso-pic}

\usepackage{newtxtext,newtxmath}

\title[Optical Photometric Variability of Quasars]{Optical Variability of Quasars with 20-Year Photometric Light Curves}

\author[Stone et al.]{Zachary Stone$^{1}$\thanks{E-mail: stone28@illinois.edu (ZS)},
Yue Shen$^{1,2}$,
Colin J. Burke$^{1,3}$,
Yu-Ching Chen$^{1,3}$,
Qian Yang$^{4,1}$,
Xin Liu$^{1,2}$,
\newauthor
R.~A.~Gruendl$^{1,3}$,
M.~Adam\'ow$^{3}$,
F.~Andrade-Oliveira$^{5,6}$,
J.~Annis$^{7}$,
D.~Bacon$^{8}$,
E.~Bertin$^{9,10}$,
\newauthor
S.~Bocquet$^{11}$,
D.~Brooks$^{12}$,
D.~L.~Burke$^{13,14}$,
A.~Carnero~Rosell$^{6}$,
M.~Carrasco~Kind$^{1,3}$,
J.~Carretero$^{15,16}$,
\newauthor
L.~N.~da Costa$^{6,17}$,
M.~E.~S.~Pereira$^{19}$,
J.~De~Vicente$^{20}$,
S.~Desai$^{21}$,
H.~T.~Diehl$^{7}$,
P.~Doel$^{12}$,
I.~Ferrero$^{22}$,
\newauthor
D.~N.~Friedel$^{3}$,
J.~Frieman$^{7,13}$,
J.~Garc\'ia-Bellido$^{23}$,
E.~Gaztanaga$^{24,25}$,
D.~Gruen$^{26}$,
G.~Gutierrez$^{7}$,
\newauthor
S.~R.~Hinton$^{27}$,
D.~L.~Hollowood$^{28}$,
K.~Honscheid$^{29,30}$,
D.~J.~James$^{4}$,
K.~Kuehn$^{31,32}$,
N.~Kuropatkin$^{7}$,
\newauthor
C.~Lidman$^{33,34}$,
M.~A.~G.~Maia$^{6,17}$,
F.~Menanteau$^{1,3}$,
R.~Miquel$^{35,36}$,
R.~Morgan$^{37}$,
F.~Paz-Chinch\'{o}n$^{3,38}$,
\newauthor
A.~Pieres$^{6,17}$,
A.~A.~Plazas~Malag\'on$^{39}$,
M.~Rodriguez-Monroy$^{20}$,
E.~Sanchez$^{20}$,
V.~Scarpine$^{7}$,
\newauthor
S.~Serrano$^{24,25}$,
I.~Sevilla-Noarbe$^{20}$,
M.~Smith$^{40}$,
E.~Suchyta$^{41}$,
M.~E.~C.~Swanson$^{42}$,
G.~Tarl\'e$^{18}$,
\newauthor
and C.~To$^{29}$
(DES Collaboration)
\newauthor
\\
\textup{\normalsize The authors' affiliations are shown at the end of this paper.}
}

\date{Accepted XXX. Received YYY; in original form ZZZ}

\pubyear{2021}

\begin{document}
\label{firstpage}
\pagerange{\pageref{firstpage}--\pageref{lastpage}}
\maketitle

\begin{abstract}
We study the optical $gri$ photometric variability of a sample of 190 quasars within the SDSS Stripe 82 region that have long-term photometric coverage during $\sim 1998-2020$ with SDSS, PanSTARRS-1, the Dark Energy Survey, and dedicated follow-up monitoring with Blanco 4m/DECam. With on average $\sim 200$ nightly epochs per quasar per filter band, we improve the parameter constraints from a Damped Random Walk (DRW) model fit to the light curves over previous studies with 10--15 yr baselines and $\lesssim 100$ epochs. We find that the average damping timescale $\tau_{\rm DRW}$ continues to rise with increased baseline, reaching a median value of $\sim 750$\,days ($g$ band) in the rest-frame of these quasars using the 20-yr light curves. Some quasars may have gradual, long-term trends in their light curves, suggesting that either the DRW fit requires very long baselines to converge, or that the underlying variability is more complex than a single DRW process for these quasars. Using a subset of quasars with better-constrained $\tau_{\rm DRW}$ (less than 20\% of the baseline), we confirm a weak wavelength dependence of $\tau_{\rm DRW}\propto \lambda^{0.51\pm0.20}$. We further quantify optical variability of these quasars over days to decades timescales using structure function (SF) and power spectrum density (PSD) analyses. The SF and PSD measurements qualitatively confirm the measured (hundreds of days) damping timescales from the DRW fits. However, the ensemble PSD is steeper than that of a DRW on timescales less than $\sim$ a month for these luminous quasars, and this second break point correlates with the longer DRW damping timescale. 
\end{abstract}




\section{Introduction} \label{sec:Intro}


The optical photometric (continuum) variability of quasars encodes critical information about physical processes within the accretion disk of a rapidly accreting supermassive black hole (SMBH) that primarily emits in the rest-frame UV through optical. There has been significant progress in the past few decades in quantifying the observed optical variability of quasars with increasing sample sizes and light curve quality \citep[e.g.,][]{Giveon_etal_1999,Hawkins_2002,VandenBerk_etal_2004,deVries_etal_2005,Sesar_etal_2006,Bauer_etal_2009,MacLeod_etal_2010,MacLeod_etal_2012,Sun_etal_2014,Morganson_etal_2014,Kasliwal_etal_2015,Chen_Wang_2015,Simm_etal_2016,Caplar_etal_2017,Smith_etal_2018,Sanchez-Saez_etal_2018,Li_etal_2018,DeCicco_etal_2019,Luo_etal_2020,Tachibana_etal_2020,Laurenti_etal_2020,Xin_etal_2020,Suberlak_etal_2021}. However, the nature of optical variability of quasars is still poorly understood \citep[e.g.,][]{Ulrich_etal_1997,Padovani_etal_2017}. 

Quasars are observed to vary stochastically over a broad range of timescales and wavelengths. In the rest-frame UV-optical, quasar variability amplitude increases with timescales and decreases with wavelength \citep[e.g.,][]{VandenBerk_etal_2004}, and is observed to anti-correlate with luminosity and the Eddington ratio of the quasar \citep[e.g.,][]{Ai_etal_2010,Rumbaugh_etal_2018}. On months to years timescales, quasar optical variability typically saturates at the $\sim 10-20\%$ level. Traditionally, the characterization of quasar variability has been carried out with the structure function (SF) or power spectrum density (PSD) measurements, which quantify the variability level as a function of timescale (or frequency). 

It has become increasingly popular in recent years to model quasar light curves in the time domain with stochastic processes \citep[e.g.,][]{Kelly_etal_2009,Kozlowski_etal_2010,Kelly_etal_2014}. This approach addresses concerns of sampling and windowing effects that come with time series analyses in the frequency domain, which are particularly relevant for ground-based quasar light curves. The Damped Random Walk (DRW) model has emerged as the simplest Gaussian random process model that can fit the optical light curves of quasars reasonably well \citep[e.g.,][]{Kelly_etal_2009,Kozlowski_etal_2010,MacLeod_etal_2010}. Deviations from the DRW model have been reported \citep[e.g.,][]{Mushotzky_etal_2011,Zu_etal_2013,Kasliwal_etal_2015,Guo_etal_2017}, although some of these claims are likely impacted by the limited duration of the light curve in the DRW fit \citep[e.g.,][]{Kozlowski_2017}. More complex stochastic process models, such as the continuous auto-regressive moving-average \citep[CARMA;][]{Kelly_etal_2014} models, can accommodate a broader range of PSD shapes, and improve the fits provided that the light curve quality is sufficiently high. In general the CARMA models do not have to be solutions to the stochastic differential equation driven by a Gaussian process (i.e., a Wiener process). However, for CARMA processes that are Gaussian, the model parameters can be estimated using efficient implementations of Gaussian process regression \citep[e.g.,][]{Celerite,Yu_Richards_2022}. In this work we focus on CARMA processes that are Gaussian. 

\begin{figure}
\centering
\includegraphics[width=.44\textwidth]{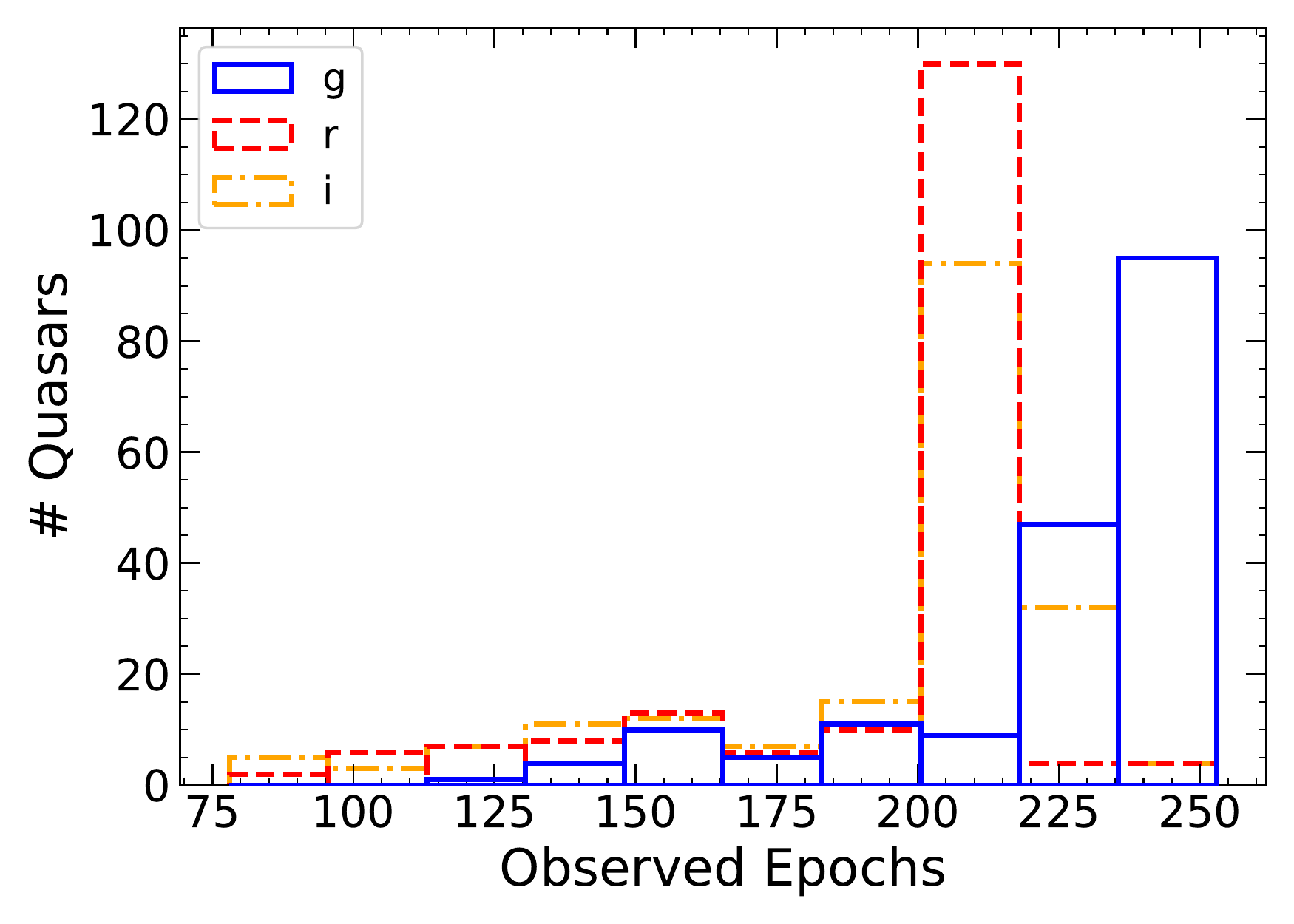}
\caption{The number of nightly-coadded epochs observed for the 190 quasars in our sample combining SDSS, PS1 and DES (+DECam) data. The \emph{gri} light curves have a median of [205, 209, 209] epochs for our quasars.}\label{fig:EpochsPerTarget}
\end{figure}

In the DRW model, the PSD is described by a $f^{-2}$ power-law at the high-frequency end, transitioning to a white noise at the low-frequency end. The transition frequency $f_0$ corresponds to the damping timescale $\tau_{\rm DRW}$ as $f_{0}=1/(2\pi\tau_{\rm DRW})$. The damping timescale thus describes a characteristic timescale of the optical variability. Earlier studies of quasar variability already hinted at such a characteristic variability timescale and its possible dependence on the physical properties of quasars such as the black hole mass \citep[e.g.,][]{Collier_Peterson_2001,Kelly_etal_2009}, but the exact form of the dependence is debated \citep[e.g.,][]{MacLeod_etal_2010,Simm_etal_2016}. Recently, \citet{Burke_etal_2021} measured the damping timescales using the DRW model for a sample of Active Galactic Nuclei (AGNs) with high-quality optical light curves over a large dynamic range in black hole mass. They found a strong positive correlation between $\tau_{\rm DRW}$ and black hole mass, which extends to the stellar mass regime with optical variability measured for nova-like accreting white dwarfs \citep[][]{Scaringi_etal_2015}. Compared with higher-order Gaussian process models, the DRW model contains a single characteristic timescale, making it easier to interpret the variability and to connect variability to the underlying physical processes \citep[e.g.,][]{Burke_etal_2021,Sun_etal_2020}. 

However, as \citet{Kozlowski_2017} pointed out, in order to constrain the damping timescale $\tau_{\rm DRW}$ when fitting the light curve with a DRW model, it is important that the duration of the light curve is substantially longer than $\tau_{\rm DRW}$. For light curves shorter than a few times $\tau_{\rm DRW}$, the measured $\tau_{\rm DRW}$ can be systematically biased low and saturated around 20--40\% of the light curve duration, with elevated scatter in the measurements. Many of the DRW fits to SDSS Stripe 82 quasars in \citet{MacLeod_etal_2010} do not pass this duration test, and their reported $\tau_{\rm DRW}$ values may be underestimated. \citet{Suberlak_etal_2021} extended the Stripe 82 light curves by another 5 years using the PanSTARRS-1 (PS1) data \citep{Chambers2016}, which alleviated this problem. But many of the updated $\tau_{\rm DRW}$ measurements are still not short enough compared with the baseline. In addition, the number of PS1 epochs is small compared with the SDSS data, and the DRW fits are likely still dominated by the SDSS light curves.

The main purpose of this work is to study optical continuum variability of a sample of quasars with a more extended 20-yr baseline. This sample represents one of the best-quality light curve data sets to study quasar variability, with hundreds of epochs from SDSS, PS1 and the high-cadence/high-S/N monitoring from the Dark Energy Survey, as well as our dedicated follow-up photometric monitoring with DECam on the CTIO-4m Blanco telescope. We will improve the DRW measurements using these extended light curves and quantify the general optical variability properties with SF and PSD analyses. 

This paper is organized as follows. In \S\ref{sec:Data} we describe the sample and the photometric light curve data. In \S\ref{sec:Results} we present our variability measurements, with the technical details provided in Appendix \ref{sec:appendix}. We discuss the implications of our results in \S\ref{sec:Disc} and conclude in \S\ref{sec:Conc}. Throughout this paper we adopt a flat $\Lambda$CDM cosmology with cosmological parameters $\Omega_{M,0}=0.3$ ($\Omega_{\Lambda,0}=0.7$) and $H_0=70\,{\rm kms^{-1}Mpc^{-1}}$. By default all timescales are in the rest-frame of the quasar unless otherwise specified; in cases where ambiguity may arise in the context we use subscripts ``$_{\rm rest}$'' and ``$_{\rm obs}$'' to explicitly refer to rest-frame and observed-frame timescales. 

\section{Data} \label{sec:Data}

\begin{figure}
\includegraphics[width=.48\textwidth]{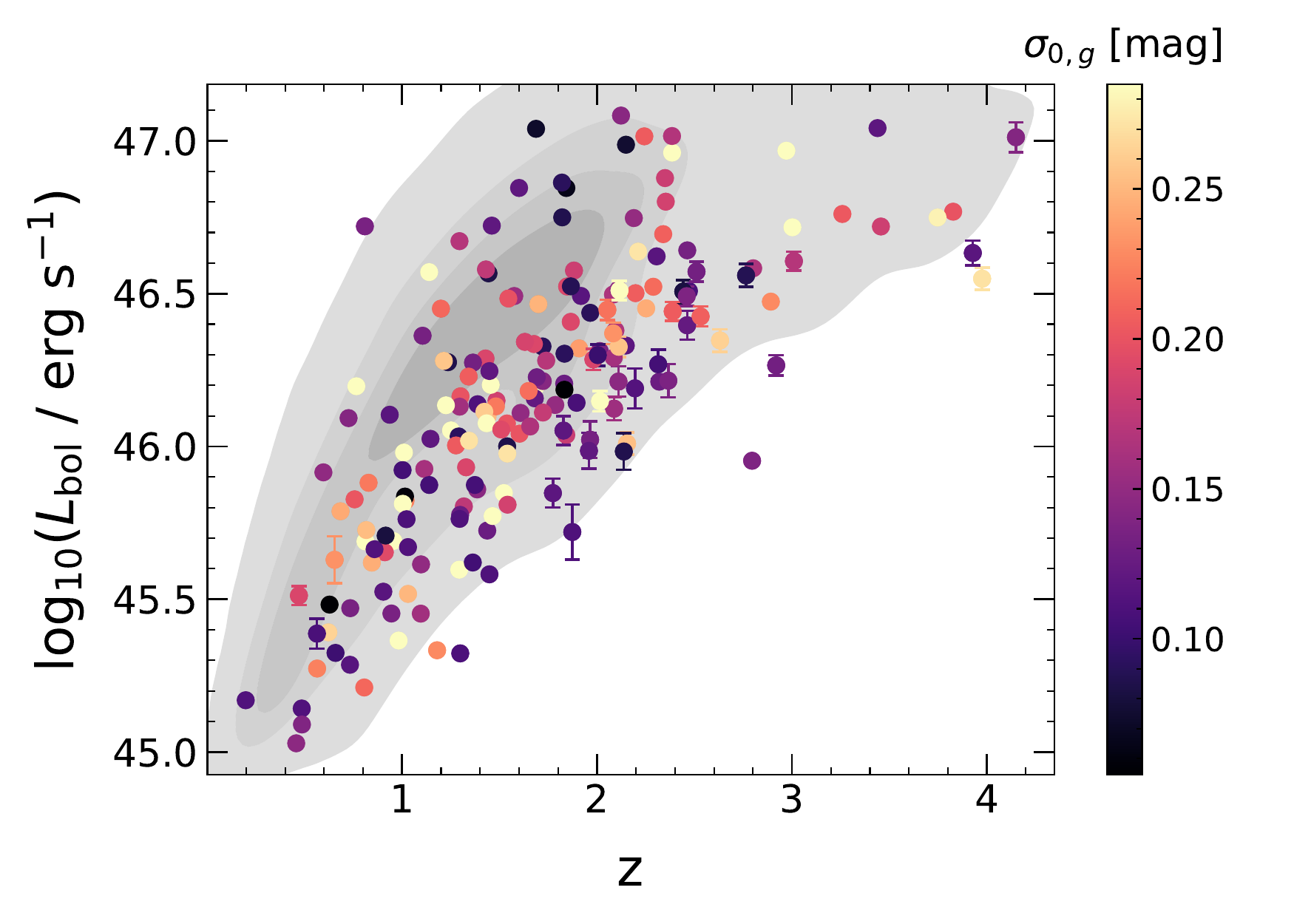}
\caption{The distribution of 190 SDSS S82 quasars in our sample in the bolometric luminosity versus redshift plane. The individual targets are color-coded by their intrinsic rms variability in the g-band ($\sigma_{0,g}$), calculated using a maximum-likelihood approach described in \citet{Shen_2019}. The gray contours behind the data points represent the distribution of $L_{\rm bol}$ and $z$ from $\sim 100,000$ SDSS DR7 quasars \citep{Shen_etal_2011}, which are on average brighter than SDSS quasars selected in the S82 region.}\label{fig:LvZ}
\end{figure}

To study optical quasar variability with long-term light curves, we utilize quasars identified in the SDSS Stripe 82 region (S82), a nearly 300 deg$^2$ stripe along the celestial equator, imaged by SDSS from $\sim 1998$ to 2007. S82 was repeatedly observed to find supernova, being one of the most frequently observed areas in SDSS. Each target within S82 was repeatedly observed for an average of 60 epochs, albeit aperiodically and with large time gaps, as the observing window spanned 2-3 months each year. SDSS photometry has five bandpasses (\emph{(ugriz)$_{\rm SDSS}$}) available for each quasar, allowing for the study of variability as a function of wavelength. The SDSS light curves in S82 provide an initial 10-year baseline for quasar variability studies \citep[e.g.,][]{MacLeod_etal_2010}. To extend this baseline, we use data from PS1 \citep{Chambers2016} spanning nearly 5 years during 2010-2014. PS1 imaged the sky in the \emph{(grizy)$_{\rm PS1}$} bandpasses with $\sim 2$ epochs per year in its wide-area survey. The combined SDSS+PS1 light curves for S82 quasars have a baseline of $\sim 15$\,yrs, and were used to study quasar variability in \citet{Suberlak_etal_2021} to improve the DRW fits. However, there were only a handful of PS1 epochs, and the DRW fits were potentially dominated by the SDSS data.  

To extend our baseline further, we use data from the DES survey during 2013-2019, which imaged the sky in the \emph{(grizy)$_{\rm DES}$} bandpasses. In particular, among the repeatedly observed DES Transient Survey (Deep) Fields~\citep{hartley_2022}, there
were two in the S82 region (SN-S1 and SN-S2, centered at J2000 coordinates 02:51:16.8$+$00:00:00.0 and 02:44:46.7$-$00:59:18.2, respectively), each with 2.7\,${\rm deg}^2$ area, with $>100$ epochs in each band over six years. The light curves in different bands have similar cadences, but are not necessarily simultaneous (i.e., on the same nights). After DES completed its wide field survey in 2019, we continued to monitor these two S82 DES-deep fields with a dedicated long-term program (2019-2024) using the DECam imager on the CTIO-4m telescope (NOAO program 2019B-0219; PI: X. Liu) to extend the baseline further in 3 bands (\emph{(gri)$_{\rm DES}$}). 

\begin{figure}
\includegraphics[width=.48\textwidth]{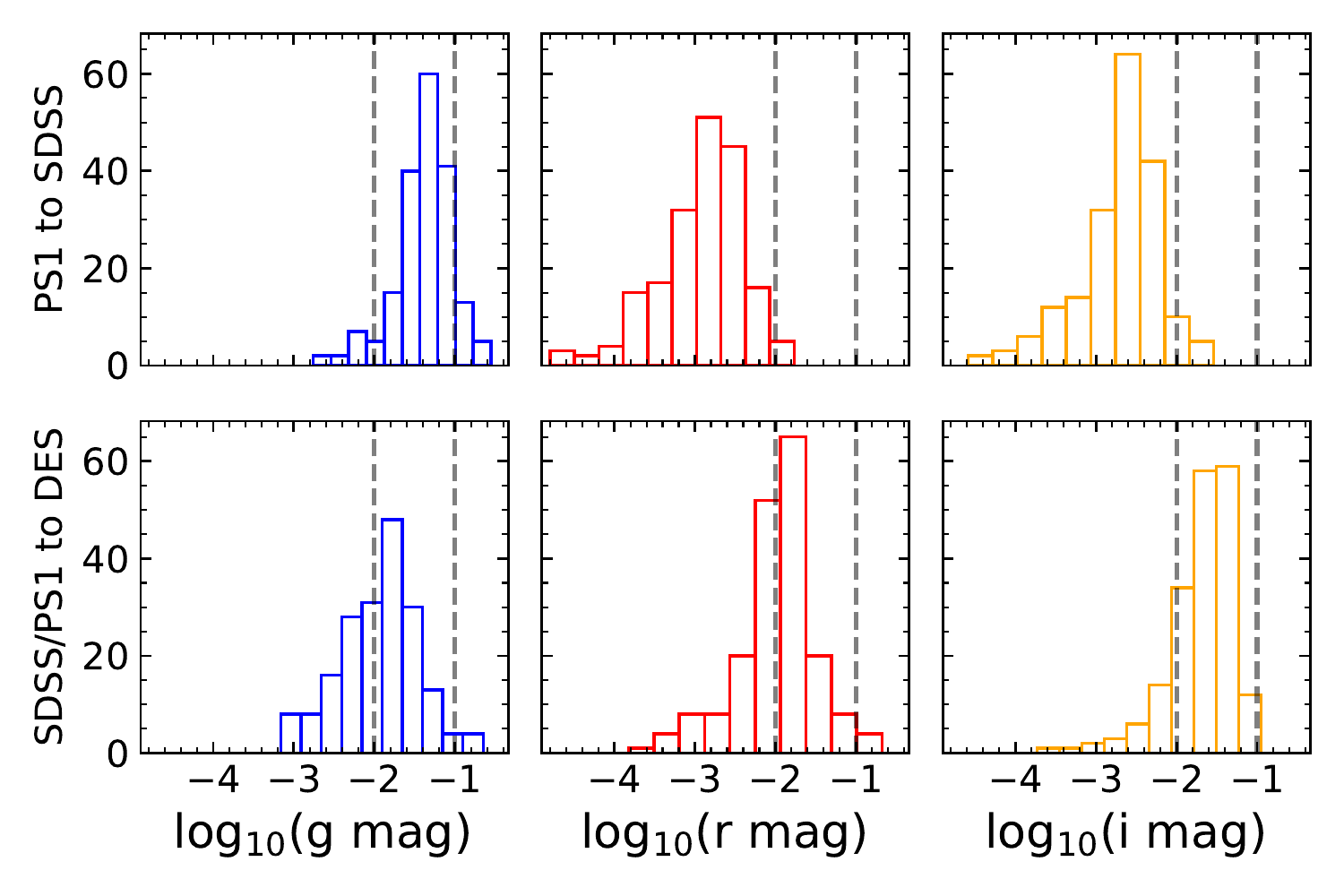}
\caption{Histograms of the photometric offsets used for each target in each survey. The top row represents offsets from PS1 bands to SDSS bands, and the bottom row represents offsets from the combined PS1/SDSS bands to DES bands. The three columns represent the \emph{gri} bands in corresponding order from left to right. The dashed lines represent 0.01 and 0.1 mag corrections for each band.}\label{fig:Offsets}
\end{figure}

In this work we use the combined light curve data from SDSS, PS1, DES and DECam imaging for 190 spectroscopically confirmed quasars in SDSS that are within the two DES-deep fields in S82 (Fig.~\ref{fig:EpochsPerTarget} and Fig.~\ref{fig:LvZ}). These quasars are all within the SDSS DR7 quasar catalog, with derived physical properties such as bolometric luminosities and black hole masses from \citet{Shen_etal_2011}. Our combined baseline is $\sim 20$\,yrs, enabling a detailed quasar variability study over decades-long timescales. The inclusion of the DES and DECam imaging is of critical importance: it not only extends the baseline by another $6$ years to improve the constraints on the damping timescale, but also provides a large number of high-S/N epochs to sample days to years timescales and to ensure the DRW fits are not dominated by the SDSS epochs. 






All of these quasars have observations in the \emph{gri} bands for all surveys, so we focus on these three bands for multi-wavelength variability. Although $z$-band data are also available across most of these surveys, the variability amplitude in this red band is lower and host contamination would be more significant, thus complicating the quasar variability measurements. We model the light curves in each band separately, instead of fitting the multi-band light curves simultaneously as did in \citet{Hu_Tak_2020}. The latter approach may be useful to further constrain inter-band correlations of the light curves.


\begin{figure}
    \centering
    \includegraphics[width=.46\textwidth]{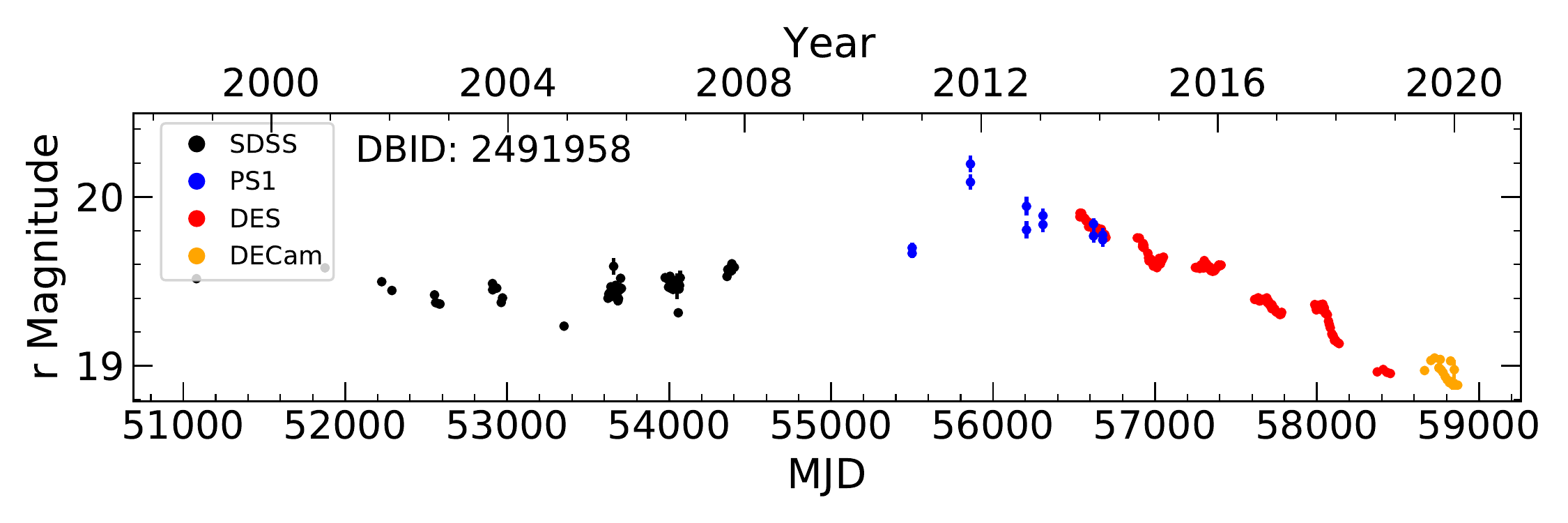}
    \includegraphics[width=.46\textwidth]{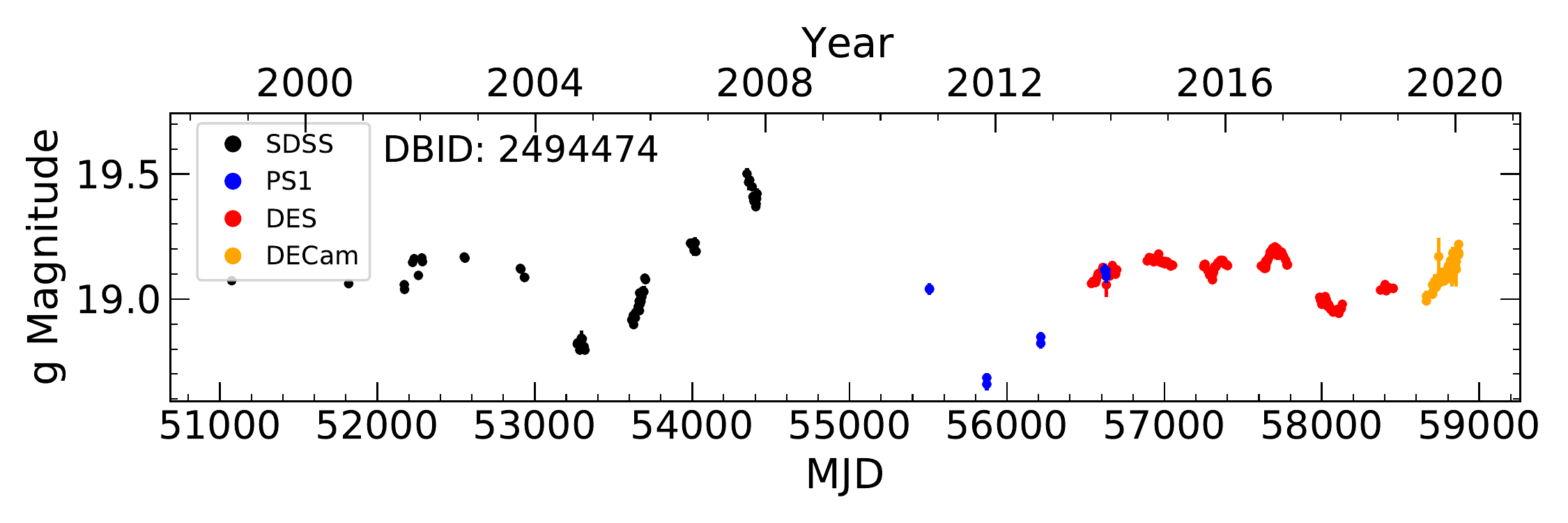}
    \includegraphics[width=.46\textwidth]{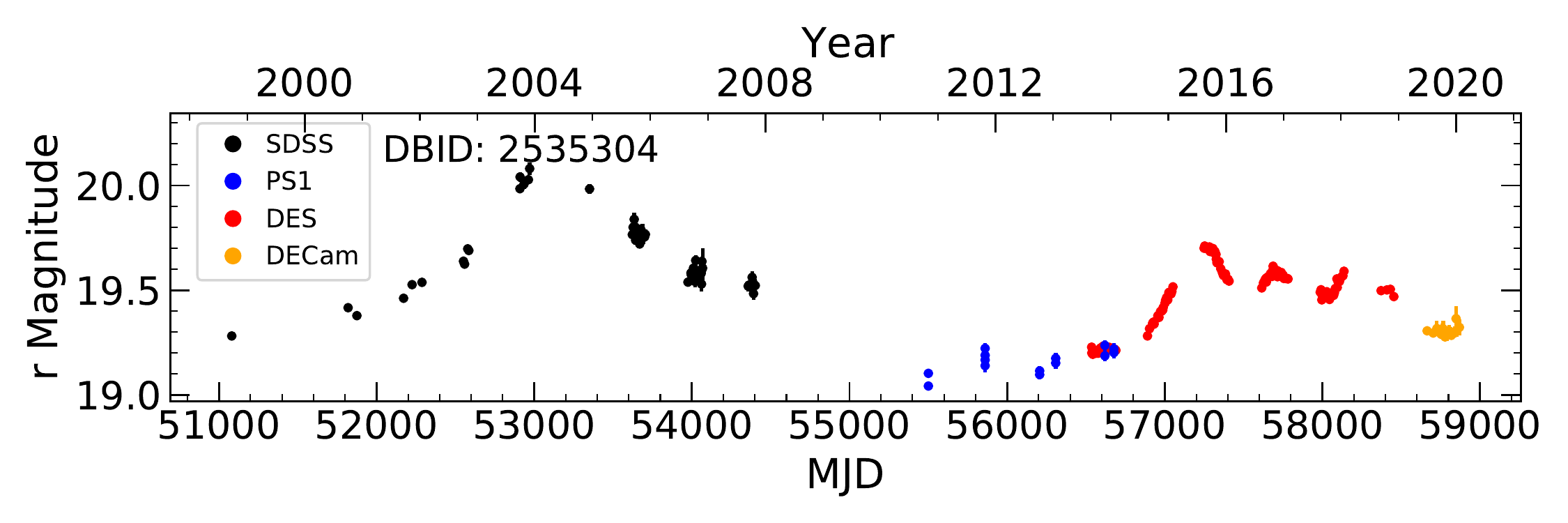}
    \caption{Example light curves from our quasar sample. These light curves are taken over a $\sim$20 year baseline, across different surveys. To adjust the observed magnitudes in a common band, we apply empirical color offsets and additional small ($\sim 0.05$ mag) offsets to merge the light curves. The data for all light curves are provided in the FITS table described in Table \ref{tab:catalogall}}
    \label{fig:lc_example}
\end{figure}

We obtain public SDSS light curve data for each of these quasars from the catalog curated in \citet{MacLeod_etal_2012}, which provides light curves for nearly 9000 SDSS S82 quasars in all five \emph{ugriz} bandpasses. We obtain public PS1 photometry for each quasar using the MAST database (\url{https://archive.stsci.edu/}), querying for all \emph{gri} bands and excluding detections with low confidence. The proprietary DES data and our dedicated DECam imaging data are processed with the same DES pipeline \citep{Morganson_etal_2018}. We use PSF magnitudes from all these surveys for our quasars. 


The filter bandpasses differ slightly between SDSS, PS1, and DES, and we apply photometric offsets to obtain merged light curves in a common bandpass for each quasar. Photometric offsets are typically constructed using colors of objects rather than magnitudes themselves, as these colors are less variable. We choose to use the mean color-based offsets described in \citet{Liu_etal_2016} to offset PS1 data into the corresponding SDSS bands, and then use the offsets described in \citet{drlica-wagner_2018} to offset both SDSS and PS1 magnitudes into the corresponding DES bandpasses. Fig.~\ref{fig:Offsets} shows that most of the corrections between surveys lie under 0.1 mag for each band. PS1 \emph{r,i} magnitudes are sufficiently similar to SDSS \emph{r,i} magnitudes so that no correction is needed, but we opt to do so for a similar processing of all bands. All other bandpasses for each survey have small offsets, with only a handful of objects with offsets up to 0.3 mag. Therefore, the use of these mean color photometric offsets is justified for our sample. 


After correcting for the zero-point offset in the same bandpass, we find that the $r$- and $i$-band light curves still display a small offset between the overlapping PS1 and DES epochs for some quasars. This additional offset is likely due to the usage of PSF magnitudes, extended host galaxy emission, seeing variations between PS1 and DES observations, as well as any residual systematics bewteen surveys. We therefore apply an additional correction ($\sim 0.05$\,mag) to manually bring the overlapping PS1 and DES epochs into agreement. We have tested w/ and w/o this minor magnitude offset between PS1 and DES and found that this detail has no effect on our variability analyses. 

We show a few representative examples of the merged light curves from SDSS+PS1+DES+DECam in Fig.~\ref{fig:lc_example}. We summarize the basic properties of our quasar sample in a FITS table along with the best-fit DRW parameters, where we compile additional properties of these quasars from the catalog in \citet{Shen_etal_2011}. The columns of this FITS table are described in Table~\ref{tab:catalogall}. We also provide all light curve data in the FITS table described in Table~\ref{tab:catalogall}.

\begin{figure*}
    \centering
    \includegraphics[width=.9\textwidth]{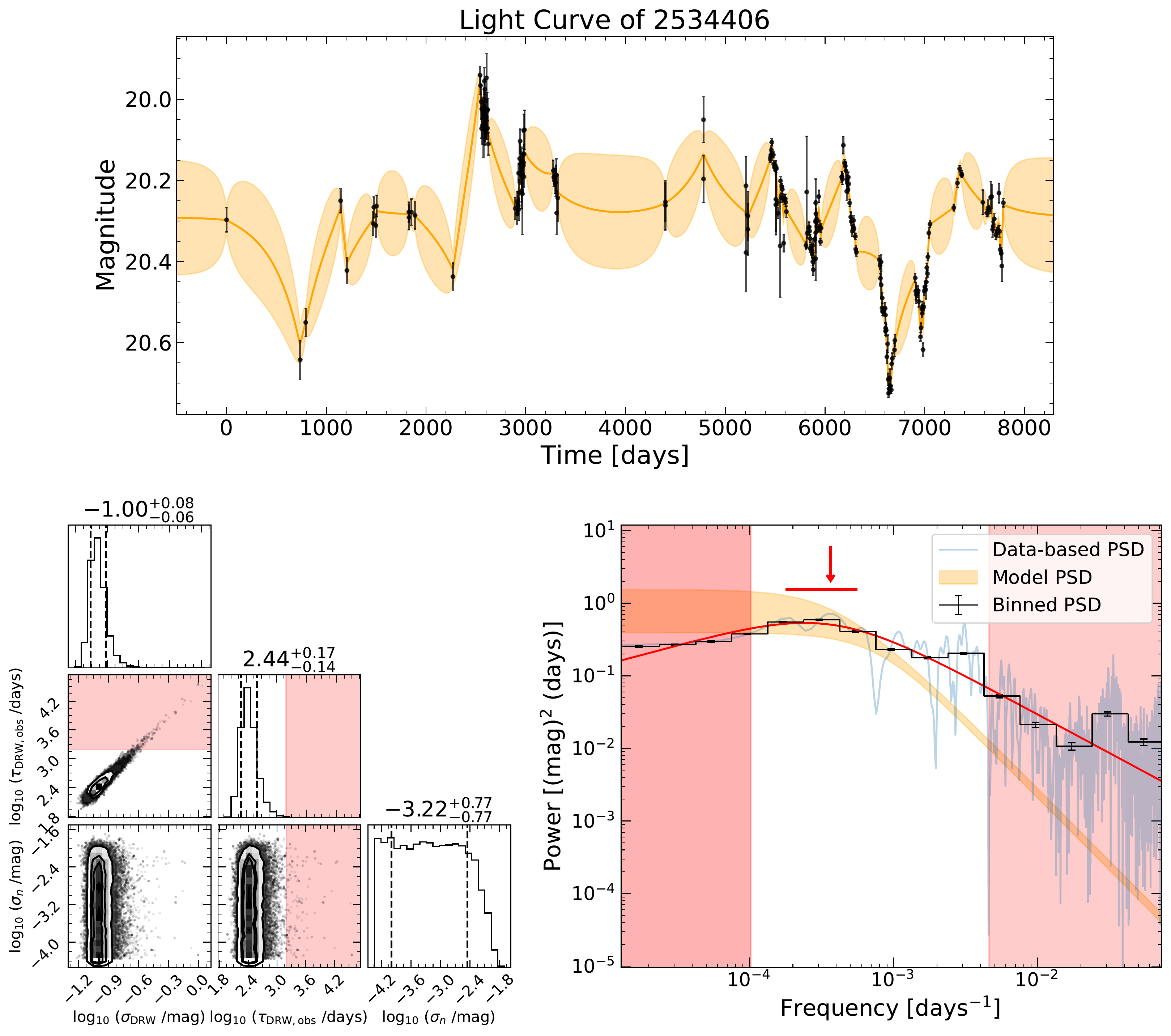}
    \caption{An example of fitting the $g$-band light curve with the DRW model using the fast Gaussian process solver {\tt Celerite} (discussed further in \ref{app:DRW}). The top panel displays the raw light curve of the object, and the predicted light curve from the DRW model using the best-fit, maximum likelihood parameters. The orange line represents the median value of the prediction, while the shaded orange region represents the area between the 1$\sigma$ uncertainty in the prediction. The plot on the lower-left displays the probability distributions of the DRW parameters fit for by {\tt Celerite}, with $\sigma_{\rm DRW}$ representing the standard deviation of long-term variability, $\tau_{\rm DRW}$ here representing the observed-frame characteristic timescale, and $\sigma_{\rm n}$ representing a noise term (also called jitter). The shaded regions in the probability distributions correspond to where $\tau_{\rm DRW, obs}$ is greater than 20\% of the baseline. The lower right plot shows the observed-frame PSD of the light curve from both the raw data and drawing from the posterior distribution of the {\tt Celerite} fit. The model PSD is shown in orange (with a band spanning the 1$\sigma$ uncertainties), the Lomb-Scargle periodogram \citep[][]{Lomb_1976,Scargle_1982} is shown in blue, and the binned Lomb-Scargle periodogram is shown in black. The binned Lomb-Scargle periodogram was also fit to a broken power law (shown as a red line), whose break frequency (and corresponding 1$\sigma$ errors) are shown with the red arrow and bar. The regions shaded red in the PSD plot correspond to regions of frequency space not sampled by the light curve (i.e. larger than the minimum cadence) as well as regions with timescales longer than 20\% of the baseline (i.e. $t > t_{\rm baseline}/5$). The difference between the Lomb-Scargle periodogram and the model PSD is caused by the difficulties of measuring the PSD accurately using the Fourier method and irregularly sampled light curves, contributions from flux uncertainties in the periodogram measurement, as well as potential deviations from a DRW model. } 
    \label{fig:taufit}
\end{figure*}


\begin{table*}
\caption{The format of the FITS table compiling the properties for our sample of 190 quasars in S82.
\label{tab:catalogall}}
\small
\begin{tabular}{llll}
\hline
$^{a}$Column Name & Format & Unit & Description \\
\hline
DBID & int64 &  & Database ID for each quasar; same as in \citet{MacLeod_etal_2012} \\
RA & float64 & deg & Right ascension of the target \\
DEC & float64 & deg & Declination of the target \\
Z & float64 &  & Redshift \\
log\_M\_BH & float64 & $\log_{10}( M_{\odot} )$ & $\log_{10}$ of the black hole mass \\
log\_M\_BH\_ERR & float64 & $\log_{10}( M_{\odot} )$ & Error in $\log_{10}$ of the black hole mass \\
log\_LBOL & float64 & $\log_{10}( {\rm erg} \ {\rm s}^{-1} )$ & $\log_{10}$ of the bolometric luminosity \\
log\_LBOL\_ERR & float64 & $\log_{10}( {\rm erg} \ {\rm s}^{-1} )$ & Error in $\log_{10}$ of the bolometric luminosity \\
log\_TAU\_OBS\_x & float64 & $\log_{10}( {\rm days} )$ & $\log_{10}( \tau_{\rm DRW} )$ in the observed-frame \\
log\_TAU\_OBS\_x\_ERR\_L & float64 & $\log_{10}( {\rm days} )$ & Lower error of $\log_{10}( \tau_{\rm DRW} )$ in the observed-frame \\
log\_TAU\_OBS\_x\_ERR\_U & float64 & $\log_{10}( {\rm days} )$ & Upper error of $\log_{10}( \tau_{\rm DRW} )$ in the observed-frame \\
log\_TAU\_REST\_x & float64 & $\log_{10}( {\rm days} )$ & $\log_{10}( \tau_{\rm DRW} )$ in the rest-frame \\
log\_TAU\_REST\_x\_ERR\_L & float64 & $\log_{10}( {\rm days} )$ & Lower error of $\log_{10}( \tau_{\rm DRW} )$ in the rest-frame \\
log\_TAU\_REST\_x\_ERR\_U & float64 & $\log_{10}( {\rm days} )$ & Upper error of $\log_{10}( \tau_{\rm DRW} )$ in the rest-frame \\
log\_SIGMA\_x & float64 & $\log_{10}( {\rm mag} )$ & $\log_{10}( \sigma_{\rm DRW} )$ \\
log\_SIGMA\_x\_ERR\_L & float64 & $\log_{10}( {\rm mag} )$ & Lower error of $\log_{10}( \sigma_{\rm DRW} )$ \\
log\_SIGMA\_x\_ERR\_U & float64 & $\log_{10}( {\rm mag} )$ & Upper error of $\log_{10}( \sigma_{\rm DRW} )$ \\
log\_JITTER\_x & float64 & $\log_{10}( {\rm mag} )$ & $\log_{10}( \sigma_{\rm n} )$ \\
log\_JITTER\_x\_ERR\_L & float64 & $\log_{10}( {\rm mag} )$ & Lower error of $\log_{10}( \sigma_{\rm n} )$ \\
log\_JITTER\_x\_ERR\_U & float64 & $\log_{10}( {\rm mag} )$ & Upper error of $\log_{10}( \sigma_{\rm n} )$ \\
SIG0\_x & float64 & mag & Intrinsic RMS variability \\
SIG0\_x\_ERR & float64 & mag & Error in intrinsic RMS variability \\
LAMBDA\_REST\_x & float64 & \AA & Rest-frame wavelength the target was observed in \\
$^{b}$SURVEY\_x & str5 &  & Imaging survey used for the observation \\
$^{b}$MJD\_x & float64 & days & MJD of the observation \\
$^{b}$MAG\_x & float64 & mag & PSF magnitude of the observation \\
$^{b}$MAG\_ERR\_x & float64 & mag & Error in the observation \\
$^{c}$OFFSET\_x & float64 & mag & Manual offset applied to the PS1 magnitudes \\
DT\_REST\_x & float64 & days & Rest-frame time lags used to construct the structure function \\
SF\_x & float64 & mag & Structure function measurements \\
SF\_x\_ERR\_L & float64 & mag & Lower error in the structure function \\
SF\_x\_ERR\_U & float64 & mag & Upper error in the structure function \\
CARMA\_P\_x & int64 &  & CARMA model p parameter \\
CARMA\_Q\_x & int64 &  & CARMA model q parameter \\
REST\_FREQ\_x & float64 & days$^{-1}$ & Rest-frame frequency \\
CARMA\_PSD\_x & float64 & (mag)$^{2}$ (days) & Median PSD constructed from the CARMA model \\
CARMA\_PSD\_x\_ERR\_L & float64 & (mag)$^{2}$ (days) & Lower error in the CARMA PSD \\
CARMA\_PSD\_x\_ERR\_U & float64 & (mag)$^{2}$ (days) & Upper error in the CARMA PSD \\
$^{d}$CARMA\_AR0\_x & float64 &  & 0$^{\rm th}$ CARMA auto-regressive parameter ($\alpha_0$) \\
$^{d}$CARMA\_AR1\_x & float64 &  & 1$^{\rm st}$ CARMA auto-regressive parameter ($\alpha_1$) \\
$^{d}$CARMA\_AR2\_x & float64 &  & 2$^{\rm nd}$ CARMA auto-regressive parameter ($\alpha_2$) \\
$^{d}$CARMA\_AR3\_x & float64 &  & 3$^{\rm rd}$ CARMA auto-regressive parameter ($\alpha_3$) \\
$^{d}$CARMA\_AR4\_x & float64 &  & 4$^{\rm th}$ CARMA auto-regressive parameter ($\alpha_4$) \\
$^{d}$CARMA\_AR5\_x & float64 &  & 5$^{\rm th}$ CARMA auto-regressive parameter ($\alpha_5$) \\
$^{d}$CARMA\_AR6\_x & float64 &  & 6$^{\rm th}$ CARMA auto-regressive parameter ($\alpha_6$) \\
$^{d}$CARMA\_AR7\_x & float64 &  & 7$^{\rm th}$ CARMA auto-regressive parameter ($\alpha_7$) \\
$^{d}$CARMA\_MA0\_x & float64 &  & 0$^{\rm th}$ CARMA moving-average parameter ($\beta_0$) \\
$^{d}$CARMA\_MA1\_x & float64 &  & 1$^{\rm st}$ CARMA moving-average parameter ($\beta_1$) \\
$^{d}$CARMA\_MA2\_x & float64 &  & 2$^{\rm nd}$ CARMA moving-average parameter ($\beta_2$) \\
$^{d}$CARMA\_MA3\_x & float64 &  & 3$^{\rm rd}$ CARMA moving-average parameter ($\beta_3$) \\
$^{d}$CARMA\_MA4\_x & float64 &  & 4$^{\rm th}$ CARMA moving-average parameter ($\beta_4$) \\
$^{d}$CARMA\_MA5\_x & float64 &  & 5$^{\rm th}$ CARMA moving-average parameter ($\beta_5$) \\
$^{d}$CARMA\_MA6\_x & float64 &  & 6$^{\rm th}$ CARMA moving-average parameter ($\beta_6$) \\
\hline
\end{tabular}\\
{\raggedright 
$^{a}$Each column labeled with ``x" is three columns, with ``x" representing the value obtained from data in the \emph{g}, \emph{r}, or \emph{i} bands. \\

$^{b}$FITS tables require that each entry in a column of data have the same length. However, each object has a different amount of epochs, making their data arrays unequal. To circumvent this, we have made the arrays corresponding to properties of the observations of the object (SURVEY, MJD, MAG, MAG$\textunderscore$ERR) the same length. This length is the number of observations for the object with the maximum number of observations in the sample. For arrays with a length less than this maximum length, we fill the arrays with NaNs or empty strings until they reach this length. \\

$^{c}$This manual offset is used to bring the PS1 and DES magnitudes into agreement in the overlapping region. Offsets were only applied to \emph{r}-band and \emph{i}-band light curves, so the ``x" here corresponds to $r$ and $i$ only. \\

$^{d}$All of the entries for the CARMA parameters are given as 3-entry arrays, consisting of the 1$\sigma$ errors (absolute values) and median value of the parameter. This array is formatted as [lower error, value, upper error]. If the CARMA model fit to the light curve data is not a high enough order to have a certain parameter, it will have an array filled with zeros. For example, if the CARMA p parameter is 3, all CARMA auto-regressive parameters greater than 3 will be [0, 0, 0] in the FITS table. \par}
\end{table*}

\begin{table*}
\caption{The format of the FITS table compiling ensemble SF and PSD measurements from subsets of our full quasar sample.
\label{tab:catalog_sf_psd}}
\small
\begin{tabular}{llll}
\hline
$^{a}$Column Name & Format & Unit & Description \\
\hline
$^{b}$Subsample & str9 &  & Description of the ensemble \\
DBIDs$\textunderscore$x & float64 &  & Database IDs of the objects included in the ensemble \\
DT$\textunderscore$REST$\textunderscore$x & float64 & days & Rest-frame time lags used to construct the structure function \\
SF$\textunderscore$x & float64 & mag & Ensemble structure function measurements \\
SF$\textunderscore$x$\textunderscore$ERR & float64 & mag & Error in structure function measurements \\
REST$\textunderscore$FREQ$\textunderscore$x & float64 & days$^{-1}$ & Rest-frame frequency \\
CARMA$\textunderscore$PSD$\textunderscore$x & float64 & (mag)$^2$ (days) & Ensemble of the median PSDs of the optimal CARMA models for each object \\
CARMA$\textunderscore$PSD$\textunderscore$x$\textunderscore$ERR$\textunderscore$L & float64 & (mag)$^2$ (days) & Lower error in the ensemble PSD \\
CARMA$\textunderscore$PSD$\textunderscore$x$\textunderscore$ERR$\textunderscore$U & float64 & (mag)$^2$ (days) & Upper error in the ensemble PSD \\
\hline
\end{tabular}\\
{\raggedright 
$^{a}$Similar to Table \ref{tab:catalogall}, all columns with names containing an ``x'' are three separate columns, where x is replaced with \emph{gri}, corresponding to values in each of the three bands. \\
$^{b}$There are four different types of ensembles described in this table in general: the total sample, the samples split by $\tau_{\rm DRW,rest}$, and the samples split in a grid by bolometric luminosity and redshift. The total subsample is labeled ``Total'', the three samples split by $\tau_{\rm DRW,rest}$ are labeled ``Tau\{i\}'' (where $i=1,2,3$), and the five samples split by luminosity and redshift are labeled ``Lz\_grid\{ij\}'' (where $i,j=1,2,3$ represents their placement on the grid).}
\end{table*}

\section{Results} \label{sec:Results}

\subsection{DRW Fits}\label{sec:drw}

We follow the standard practice in the literature to fit a DRW model to the quasar light curve \citep[e.g.,][]{Kelly_etal_2009,Kozlowski_etal_2010,MacLeod_etal_2010,Suberlak_etal_2021,Burke_etal_2021}. The details of the DRW modeling are provided in Appendix \ref{app:DRW}. The best-fit DRW parameters are compiled in the FITS catalog described in Table 1. An example DRW fit is shown in Fig.~\ref{fig:taufit}.

\begin{figure*}
    \centering
    \includegraphics[width=\textwidth]{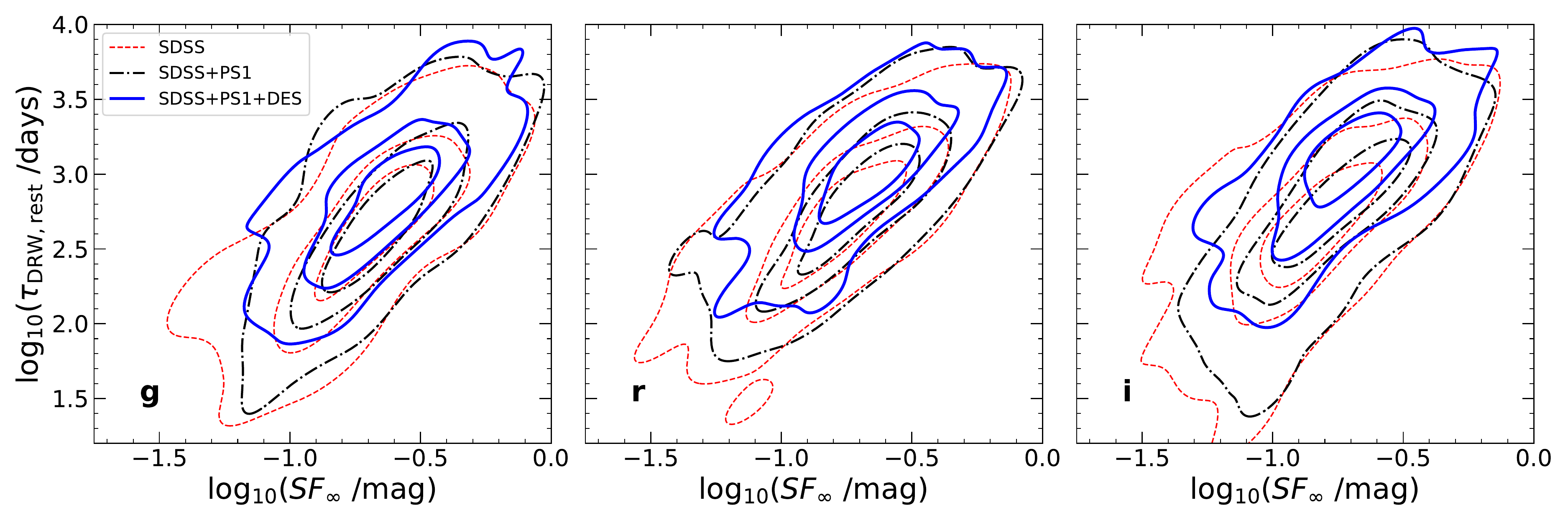}
    \caption{Contour plots showing the distribution of $SF_{\infty}$ and $\tau_{\rm DRW}$ fitted from our quasar light curve sample. There are three contours for each band, representing data fitted from light curves using only SDSS, SDSS and PS1, and all of the data. The contours for each dataset enclose [33, 66, 100]\% of the distribution respectively.}
    \label{fig:SFvTau}
\end{figure*}

In Fig.~\ref{fig:SFvTau}, we show the distribution of our sample in the $\tau_{\rm DRW}$ versus ${\rm SF}_{\infty}^2\equiv 2\sigma_{\rm DRW}^2$ plane, where $\sigma_{\rm DRW}$ is the long-term variability amplitude in the DRW model (see Appendix \ref{app:DRW}). With the SDSS-only baselines, we reproduce the results in \citet{MacLeod_etal_2010}, with a median value for $\tau_{\rm DRW, rest}$ of $\sim 540$ days in the $r$ band. Using SDSS+PS1-only baselines, however, we obtain a median value for $\tau_{\rm DRW, rest}$ of $\sim 680$ days in the $r$ band, while \citet{Suberlak_etal_2021} quoted a value of $\sim 550$ days. We attribute this discrepancy to the method of choosing the best-fit value from the DRW fit (discussed further in Appendix \ref{app:mock_lc}). By extending the baseline further with the DES+DECam data, the values of $\tau_{\rm DRW}$ and ${\rm SF}_{\infty}$ continue to rise. The median values of $\tau_{\rm DRW,rest}$ and ${\rm SF}_{\infty}$ for S82 quasars with our final baselines are $\sim 750$ days and $0.25\,$mag in $g$ band. 

\begin{figure*}
\centering
\includegraphics[width=\textwidth]{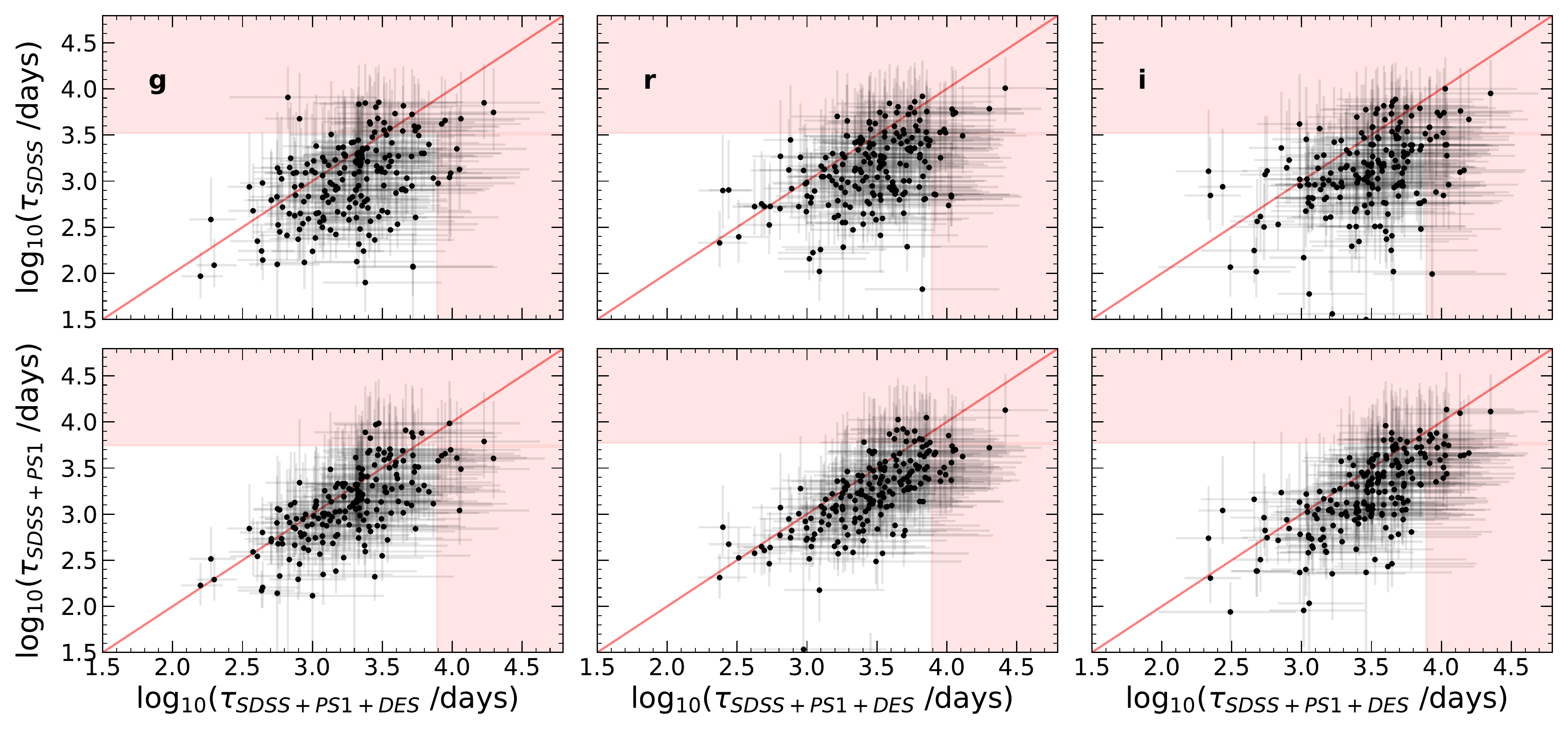}
\caption{Comparisons between $\tau_{\rm DRW, obs}$ measurements from DRW fits with different baselines. The upper panels compare the $\tau_{\rm DRW, obs}$ fitted from only using SDSS data (a $\sim$10 yr baseline) to the $\tau_{\rm DRW, obs}$ fitted from the entire 20 yr dataset for the three bands. The lower panels compare the $\tau_{\rm DRW, obs}$ fitted using data from SDSS and PS1 (a $\sim$15 yr baseline) to those fitted using the full light curves. The red shaded area indicates where $\tau_{\rm DRW, obs}$ is greater than the light curve baseline, on each respective axis. The red line running through the data shows the unity relation. }\label{fig:TauScatter}
\end{figure*}

Fig.~\ref{fig:TauScatter} compares the $\tau_{\rm DRW,obs}$ values measured with different baselines. Similar to the results shown in Fig.~\ref{fig:SFvTau}, the best-fit $\tau_{\rm DRW,obs}$ continues to increase as the baseline increases. With longer baselines and more epochs, the constraints on $\tau_{\rm DRW}$ are somewhat tighter, as demonstrated by the lower scatter of points with the SDSS+PS1 and SDSS+PS1+DES+DECam data than with the SDSS-only data in Fig.~\ref{fig:TauScatter}. However, the formal measurement uncertainties on $\tau_{\rm DRW}$ are only reduced by $\sim 10\%$ on average from the SDSS-only measurements to the SDSS+PS1+DES+DECam measurements. It is possible that the formal measurement uncertainties underestimated the true uncertainties on $\tau_{\rm DRW}$ in these studies. 




\citet{Kozlowski_2017} emphasized the importance of the length of the light curve in constraining the DRW damping timescale. The best-fit $\tau_{\rm DRW}$ could be significantly underestimated if the light curve is not long enough, as independently confirmed in other studies with simulated light curves \citep[e.g.,][]{Suberlak_etal_2021,Burke_etal_2021}. The fact that the average $\tau_{\rm DRW}$ continues to rise as the baseline increases indicates that even the 20-year baseline is probably not long enough to well constrain $\tau_{\rm DRW}$ in some S82 quasars. On the other hand, the increasing $\tau_{\rm DRW}$ as the baseline increases may be due to gradual, long-term trends in the quasar light curve (see further discussion in Appendix \ref{app:DRW}), or it is possible that these quasar light curves are more complex than a simple DRW process with only one characteristic timescale. 

Nevertheless, simulations with mock light curves have shown that the systematic bias in $\tau_{\rm DRW}$ is not significant, albeit with elevated scatter, when the measured $\tau_{\rm DRW}$ is less than $20\%$ of the baseline \citep[e.g.][]{Kozlowski_2017,Suberlak_etal_2021,Burke_etal_2021}. For example, around the $20\%$ baseline mark, the bias in the median of the measured $\tau_{\rm DRW}$ is only $\sim 0.12-0.15$\,dex from the simulations in the above studies, which is much smaller than the scatter of individual $\tau_{\rm DRW}$ measurements. Indeed, when we compare our best-fit DRW model to the ensemble SF and PSD measurements in \S\ref{sec:sf} and \S\ref{sec:psd}, we find that these DRW fits and the associated damping timescales are qualitatively correct on average.  

\begin{figure*}
    \centering
    \begin{subfigure}{}
      \centering
      \includegraphics[width=.48\linewidth]{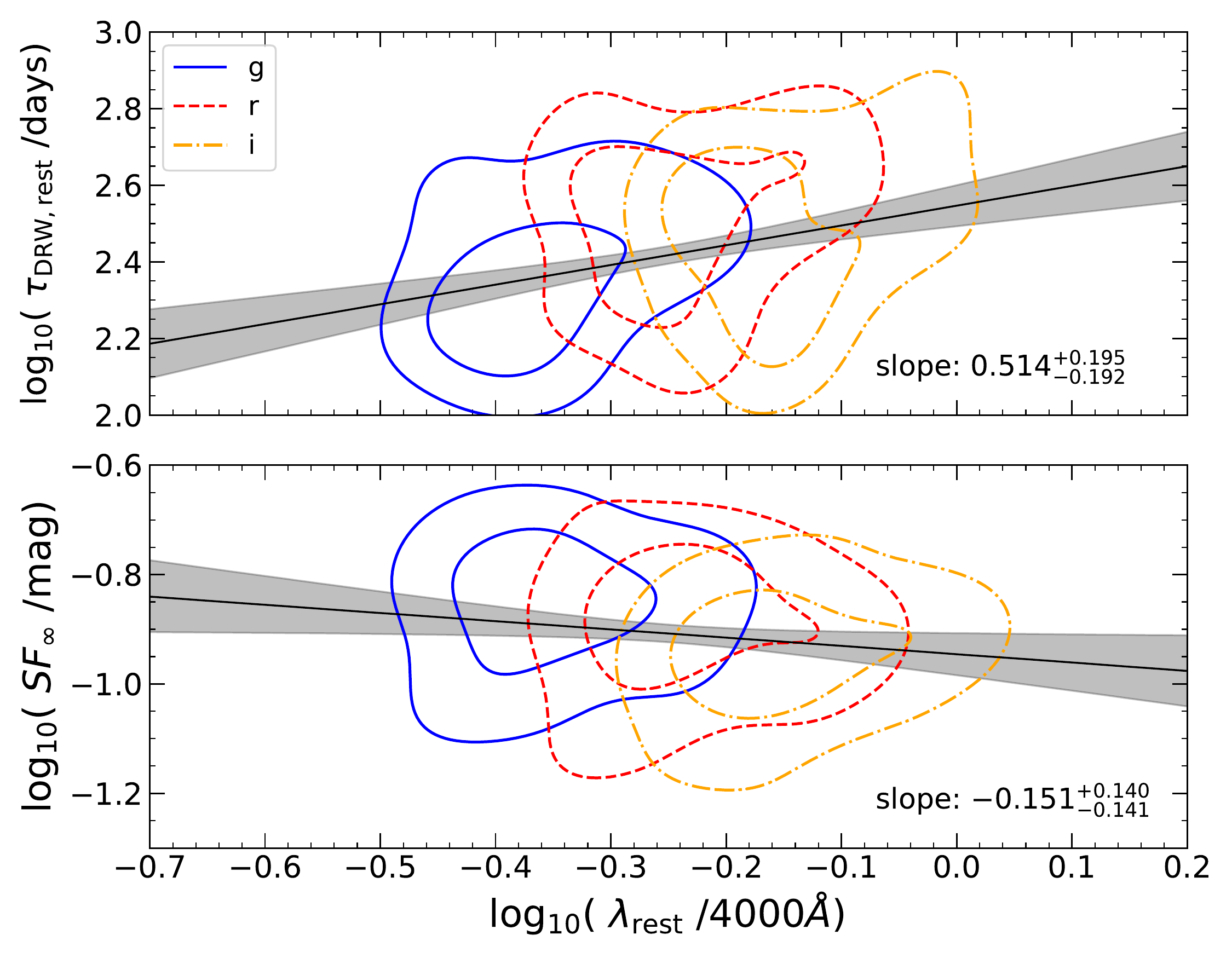}
    \end{subfigure}
    \begin{subfigure}{}
      \centering
      \includegraphics[width=.48\linewidth]{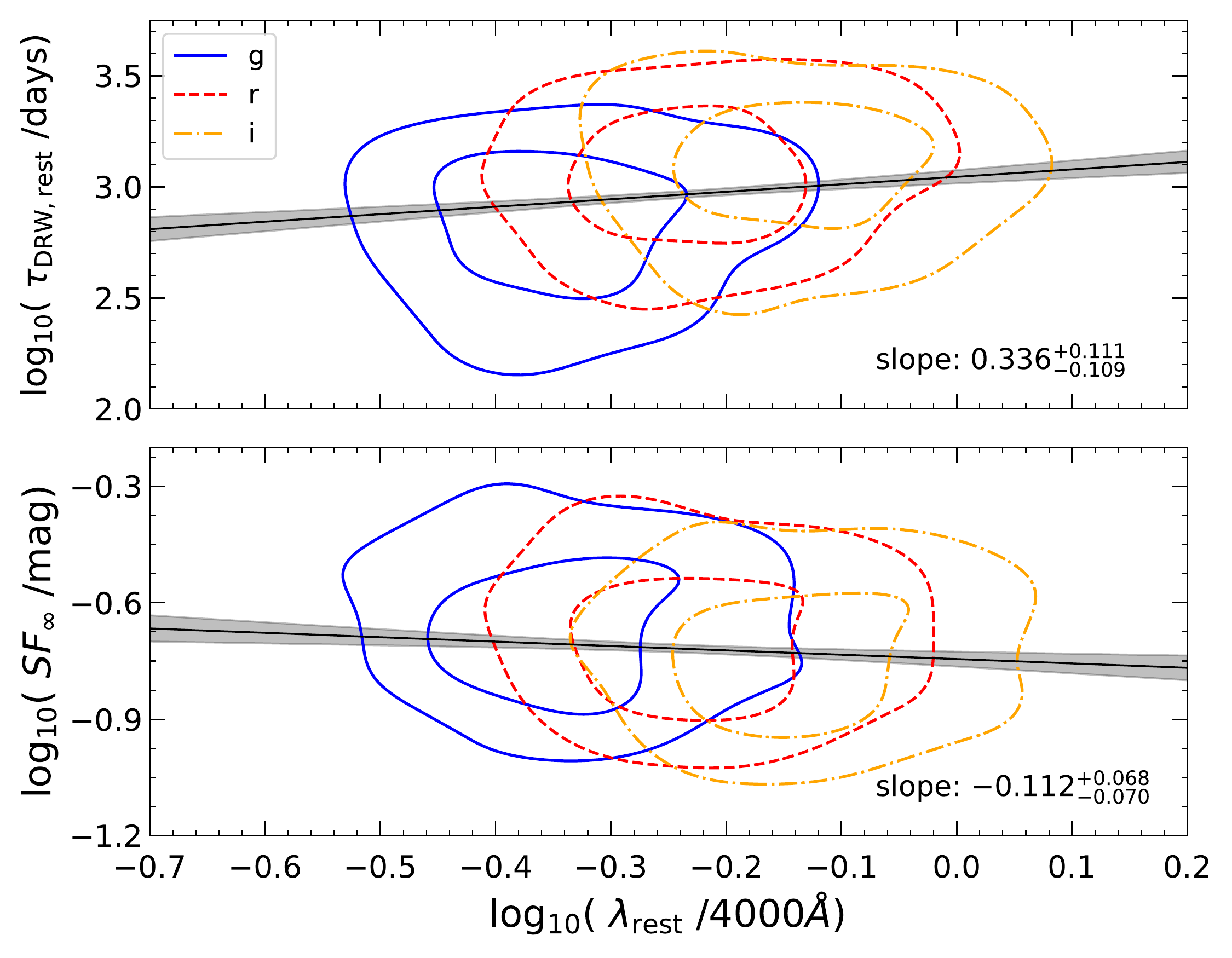}
    \end{subfigure}
    \caption{Wavelength dependences of both $\tau_{\rm DRW}$ and ${\rm SF}_{\infty}$. The left panels are for a subset of 27 quasars for which all measured $\tau_{\rm DRW, obs}$ values are less than 20\% of the final baseline. The right panels are for the full sample of 190 quasars. The contours in blue, red, and orange represent the results from \emph{g}, \emph{r}, and \emph{i} light curves, respectively, shifted to the corresponding rest-frame wavelengths of each individual quasar. The contours for each band represent 30 and 70 percent of the data. The best-fit linear regression model and $1\sigma$ uncertainties using the method described in \citet{Kelly_2007} are shown in the black line and shaded area, with the best-fit slopes marked in each panel. }
    \label{fig:WavelengthDependence}
\end{figure*}

Next, we investigate the wavelength dependence of $\tau_{\rm DRW}$ using our measurements. To reduce the impact of poorly constrained $\tau_{\rm DRW}$ values from insufficient baselines, we only use a subset of 27 quasars with measured $\tau_{\rm DRW}$ less than 20\% of the baseline, for which we consider the constraint on the damping timescale is more reliable. Using a more stringent cut on the baseline criterion would be unnecessary, and would greatly reduce our sample statistics. Fig.~\ref{fig:WavelengthDependence} (left) displays the wavelength dependences of $\tau_{\rm DRW}$ and ${\rm SF}_{\infty}$. We find a weak wavelength dependence of $\tau_{\rm DRW}\propto \lambda^{0.51\pm0.20}$, which is slightly steeper than (but formally consistent within 2$\sigma$) the one reported in \citet{MacLeod_etal_2010} based on the much shorter SDSS-only light curves $\tau_{\rm DRW}\propto \lambda^{0.17}$. On the other hand, we recover a weak anti-correlation between ${\rm SF}_{\infty}$ and wavelength, but our dynamic range in wavelength is more limited than that in \citet{MacLeod_etal_2012}, given that we only use data in $gri$ bands. These constraints on wavelength dependences are weak given the small number of quasars that pass the baseline criterion. If we use the full sample of 190 quasars instead, we find slightly different, but fully consistent results (right panel of Fig.~\ref{fig:WavelengthDependence}).

\subsection{Structure Function Analysis}\label{sec:sf}

The structure function measures the magnitude difference for pairs of epochs separated at different timescales, and is a simple and useful empirical tool to characterize the variability of quasars \citep[e.g.,][]{Collier_Peterson_2001,Kozlowski_etal_2016}. Unlike the DRW model, the SF measurements are model-independent, and provide empirical constraints on variability amplitude as a function of timescales. However, unlike the DRW and higher order CARMA modelling, the SF approach does not rigorously deal with the flux uncertainties of each epoch, and unequal flux uncertainties for long-term pairs from different surveys may complicate the SF calculation. We therefore only use these SF measurements to provide a qualitative comparison with the more rigorous DRW and CARMA PSD fits.    

For the SF analysis and the PSD analysis in \S\ref{sec:psd}, we will focus the discussion on the results using $g$-band data as we did not find significantly new information based on the $r$ and $i$-band data. However, all the individual and ensemble SF and PSD measurements for the 3 bands are compiled in Tables \ref{tab:catalogall} and \ref{tab:catalog_sf_psd}. 

\begin{figure*}
    \centering
    \includegraphics[width=\textwidth]{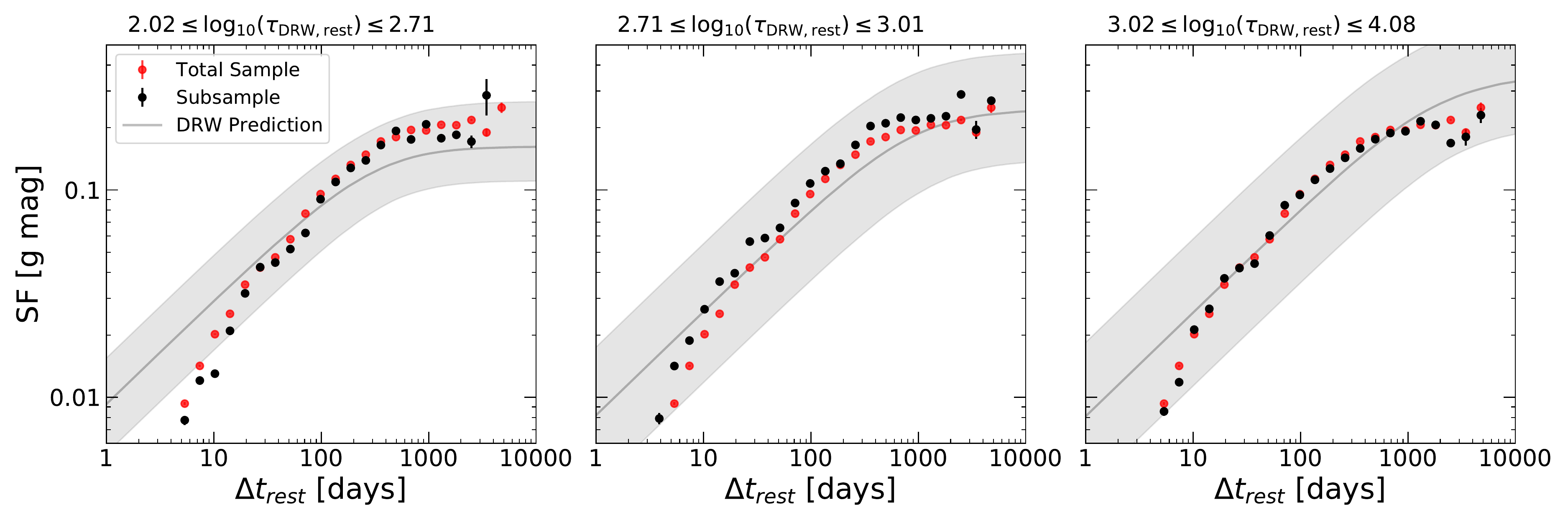}
    \caption{Ensemble structure functions for different ensembles of the 190 quasars in our sample, grouped by their fitted $\tau_{\rm DRW, rest}$. The objects are grouped such that there are an equal number of quasars in each ensemble, resulting in uneven bin widths in $\tau_{\rm DRW, rest}$. The ensemble structure function for the full sample is overlayed in red, while the structure functions for the individual subsamples are plotted in black. The predicted structure functions using the fitted $SF_{\infty}$ and $\tau_{\rm DRW}$ are plotted in gray. To obtain this DRW prediction, we sample 500 predicted DRW structure functions from each target in the ensemble, drawn from a Gaussian distribution using its best-fit DRW parameters and their uncertainties. We then combine the samples for all targets and use the median value in each $\Delta t$ bin (shown as the gray line) as the DRW-predicted structure function, and the 16$^{\rm th}$ and 84$^{\rm th}$ in each $\Delta t$ bin (colored in a gray band around the median) to construct the errors in the DRW prediction.}
    \label{fig:3tauEnsSF}
\end{figure*}

\begin{figure*}
    \centering
    \includegraphics[width=.9\textwidth]{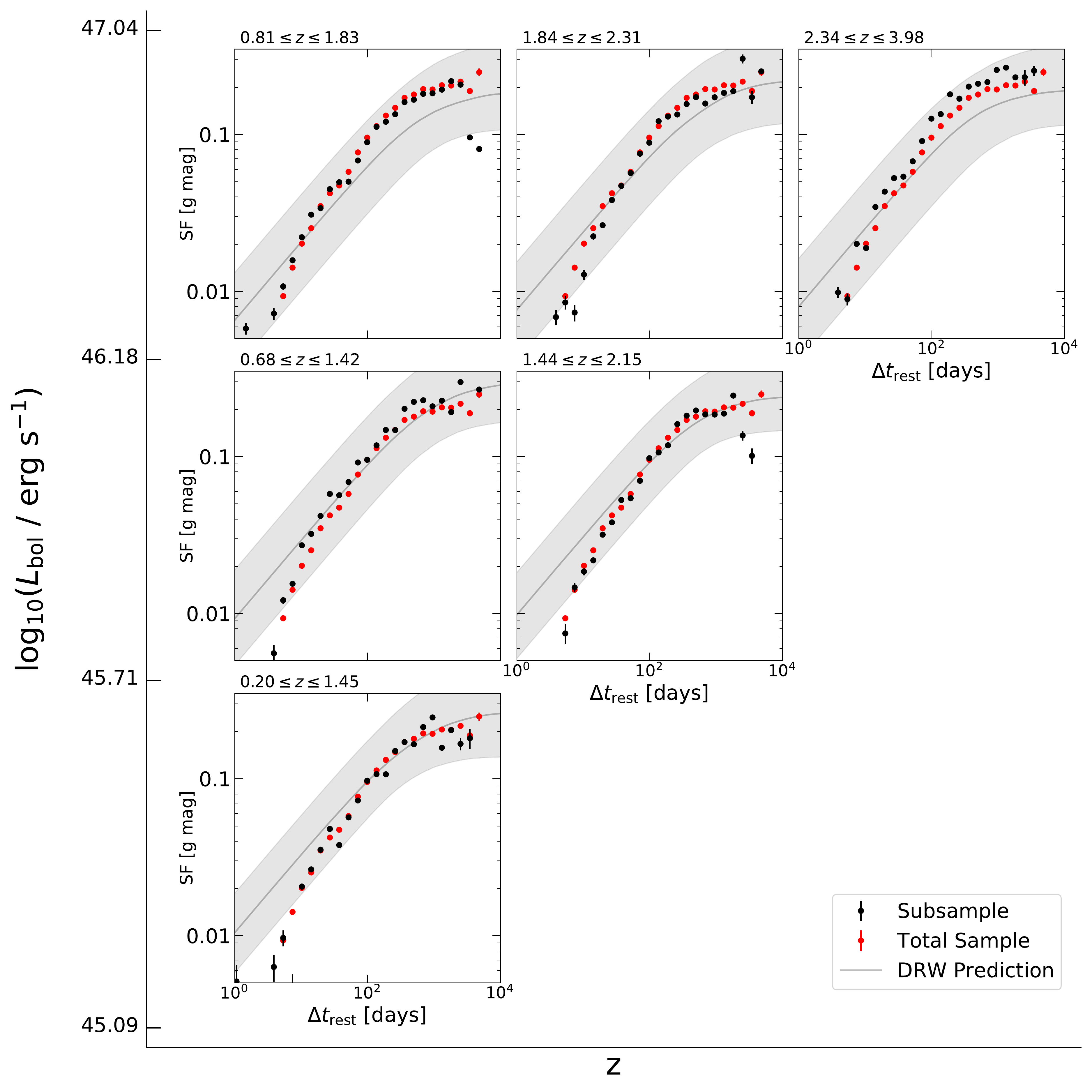}
    \caption{Ensemble $g$-band structure functions for different subsets of the full sample, grouped by their bolometric luminosity and redshift. Similar to Fig.~\ref{fig:3tauEnsSF}, we group these objects such that there is nearly an equal amount of objects in each bin. The quasars with the highest luminosities are spread over a large redshift range, which is split into three redshift bins to retain an equal number of quasars in each bin. This process was followed for the second and third luminosity bins, leaving only one redshift bin for the lowest luminosity bin. As a result, the redshift ranges are different for different luminosity bins. Each subsample contains $\sim$30 objects. The redshift ranges are listed above each subsample, and the $L_{\rm bol}$ ranges are shown on the leftmost axis, being $[45.09,45.71],[45.71,46.18]$ and, $[46.18, 47.04]$. Each subsample in a given row has the same range of $L_{\rm bol}$. We have subtracted an ``SF floor'' seen in time lags below $\sim 10$ days, to remove contamination from PSF variations and host-galaxy flux (discussed further in Appendix~\ref{app:sf}). The ensemble SF for the full sample and the DRW-prediction for each subsample are also shown for reference. We constructed the ensemble DRW-predicted structure functions in the same manner as those presented in Fig.~\ref{fig:3tauEnsSF}. }
    \label{fig:SF_grid_g}
\end{figure*}

We measure the SF for individual quasars in our sample as well as for the ensemble average. We have followed \citet{Kozlowski_etal_2016} to calculate the SF after subtracting photometric uncertainties (e.g., from flux uncertainties and additional systematics from host galaxy light and seeing variations) using close pairs separated by less than $\sim 10$ days in rest-frame. Fig.~\ref{fig:3tauEnsSF} and Fig.~\ref{fig:SF_grid_g} display the ensemble SF for different subsamples, where the full sample is divided into subsamples with approximately the same number of objects in each division (either by $\tau_{\rm DRW}$ or by $L_{\rm bol}/z$).  


Fig.~\ref{fig:3tauEnsSF} compares the ensemble SF with the median DRW model for subsets of quasars binned by the measured $\tau_{\rm DRW}$. The SF does show a flattening roughly around the location of $\tau_{\rm DRW}$ measured from the DRW fits, indicating the presence of such a damping timescale on the order of hundreds of days. 


We also recover the well known dependences of variability amplitude on wavelength and luminosity of quasars using ensemble SF measurements (data required to generate these plots are provided in Table \ref{tab:catalog_sf_psd}).


\subsection{Power Spectrum Density Analysis}\label{sec:psd}

\begin{figure}
    \centering
    \includegraphics[width=.46\textwidth]{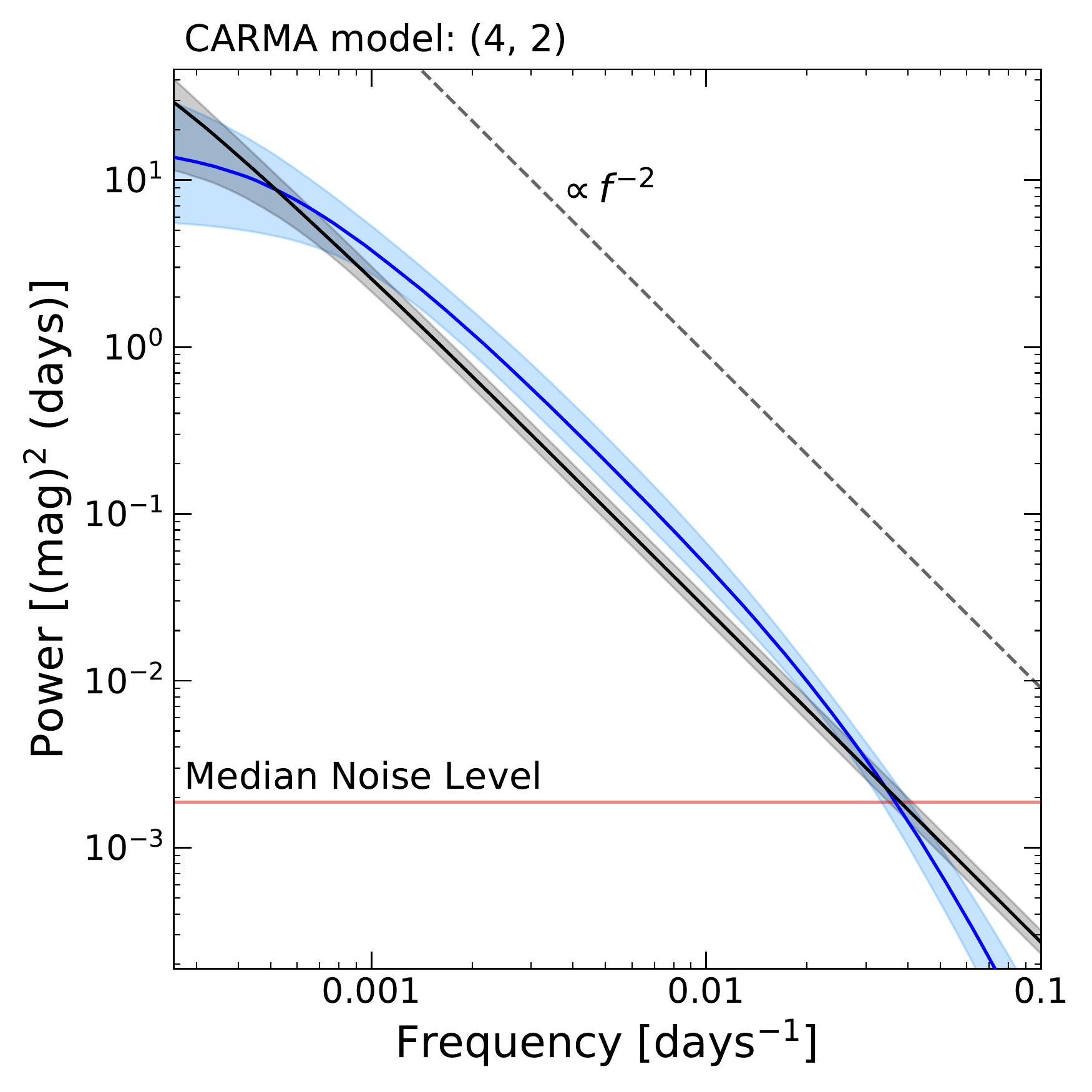}
    \caption{An example of CARMA model-fit PSDs for our quasar sample. The CARMA-predicted PSD (discussed further in \ref{app:psd}) is shown in blue, where the median from the posterior is the solid line and the shaded region encloses the 1$\sigma$ uncertainty range. The median noise level derived from the raw light curve data (2$\times$median($\Delta t$)$\times$median($\sigma_y^2$)) is shown as the red horizontal line. The gray dashed line indicates a $\propto f^{-2}$ PSD. A DRW-fit PSD for the same example light curve is shown as a black line for comparison, with the 1$\sigma$ uncertainty range shaded in gray. The CARMA-predicted PSDs for individual targets are compiled in the FITS catalog described in Table \ref{tab:catalogall}. }
    \label{fig:ExPSD}
\end{figure}

We measure the optical variability PSD using our sample and light curve data set. Because our light curves are irregularly sampled with large seasonal gaps, it is challenging to directly measure the PSD using the Fourier method, which suffers from aliasing and power leakage from windowing effects. Instead, we take advantage of the recent development of fitting Gaussian random process models to time series data and recovering the PSD \citep{Kelly_etal_2014}. Such an alternative approach is more robust in measuring the PSD with sparsely and irregularly sampled light curve data \citep[e.g.,][]{Kelly_etal_2014,Simm_etal_2016}, and properly deals with uneven measurement uncertainties in the light curve. 

Specifically, we use the {\tt CARMA\_pack} developed by \citet{Kelly_etal_2014} to find the best-fit CARMA(p,q) model to the light curve and derive the PSD, where (p,q) are the numbers of auto-regression (AR) and moving average (MA) terms, respectively. The technical details of CARMA fits are described in Appendix \ref{app:psd}. We show an example of PSD analysis in Fig.~\ref{fig:ExPSD}, and all the individual and ensemble PSDs are provided in the FITS catalogs described in Tables \ref{tab:catalogall} and \ref{tab:catalog_sf_psd}. 

\begin{figure*}
    \centering
    \includegraphics[width=\textwidth]{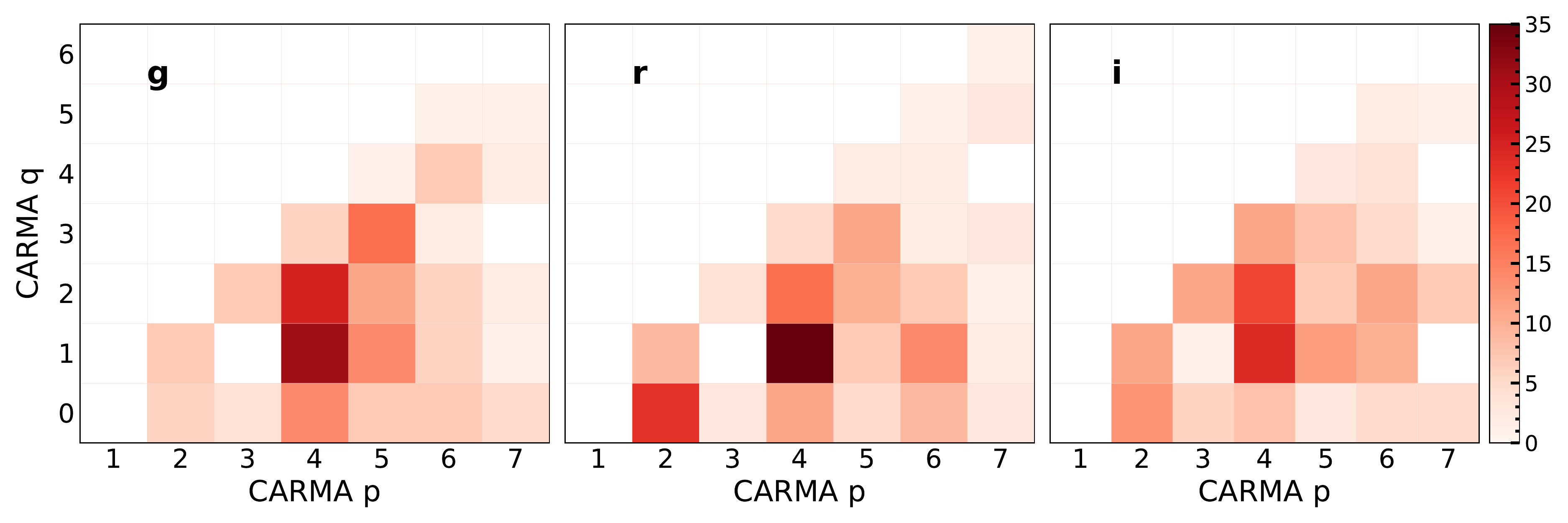}
    \caption{The distribution of best-fit CARMA $(p,q)$ parameters when fitting our quasar light curves to a generalized CARMA model (requiring $q<p$ for stationary processes). The best-fit order for the CARMA model for a given quasar light curve was chosen as the fit with the minimum value for the AICc. The AIC \citep[Akaike Information Criterion,][]{Akaike_1973} is a statistic measuring an estimate of information loss due to assuming a particular model generates a certain set of data, which can be corrected for a finite sample size to give the AICc \citep{Hurvich_1989} (discussed further in \ref{app:psd}). Darker colors indicate higher incidence. There is a tendency of clustering of quasars around $(p,q)\approx (4,2)$.}
    \label{fig:pq_dist}
\end{figure*}


We show the distributions of the best-fit values of $p$ and $q$ in Fig.~\ref{fig:pq_dist}. There is a tendency of clustering near $p\approx 4$ and $q\approx 1-2$, which may indicate the general similarity of quasar variability PSDs. However, we found that a forced CARMA(2,1) model fit produces very similar PSDs to the ones from the best-fit higher-order CARMA models. Indeed, the preference based on the model selection criterion described in Appendix \ref{app:psd} is not obvious among the higher-order CARMA models; but the preference of a higher-order CARMA model over the DRW model is often significant \citep[e.g.,][]{Kelly_etal_2014}. In particular, the CARMA(2,1) model is also known as the damped harmonic oscillator (DHO) model, which is argued to be a superior statistical description for quasar variability than the simpler DRW model \citep[e.g.,][]{Kasliwal_etal_2017,Moreno_etal_2019, Yu_etal_2022}. 

\begin{figure}
    \centering
     \includegraphics[width=0.48\textwidth]{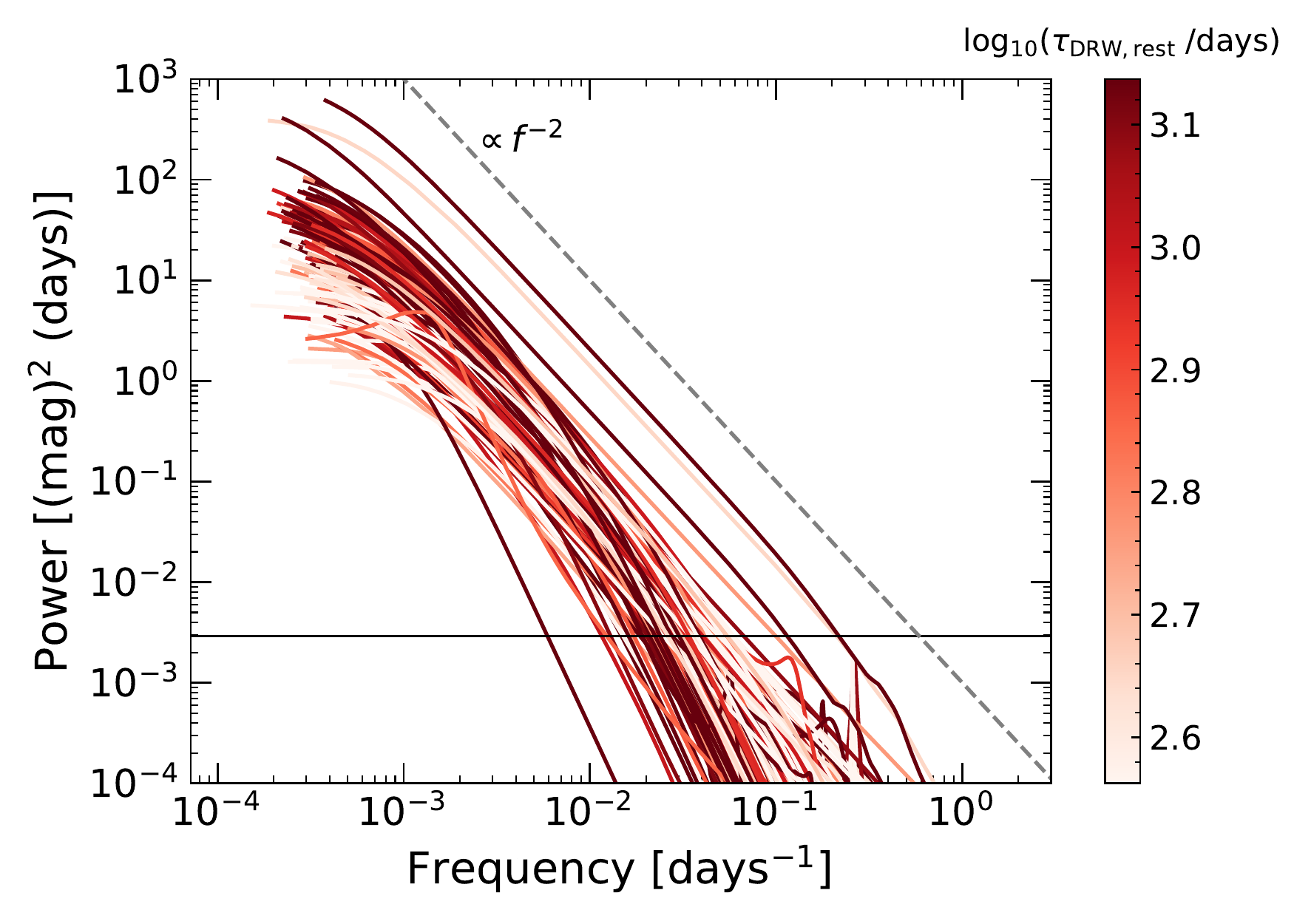}
     \includegraphics[width=0.48\textwidth]{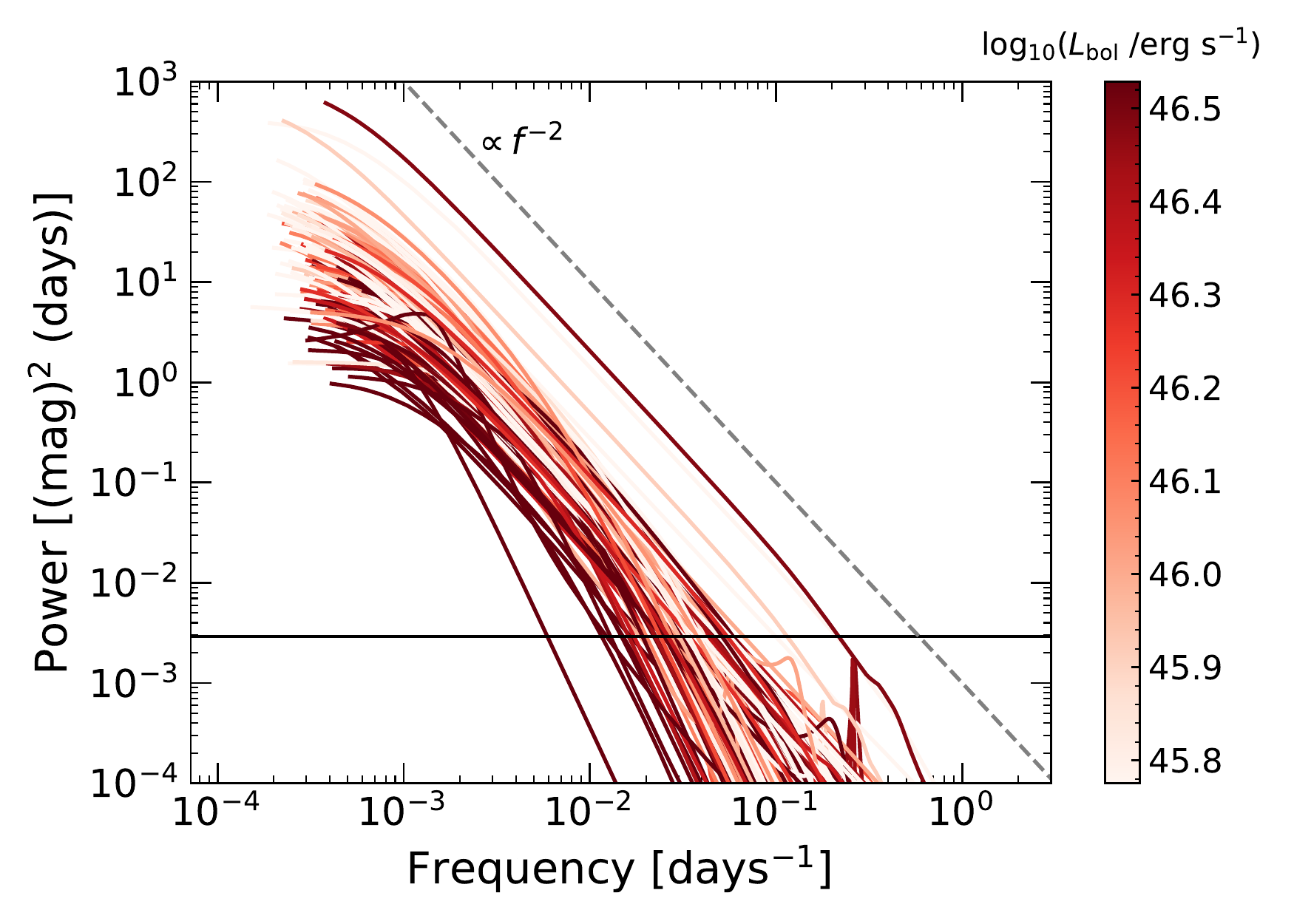}
     \includegraphics[width=0.48\textwidth]{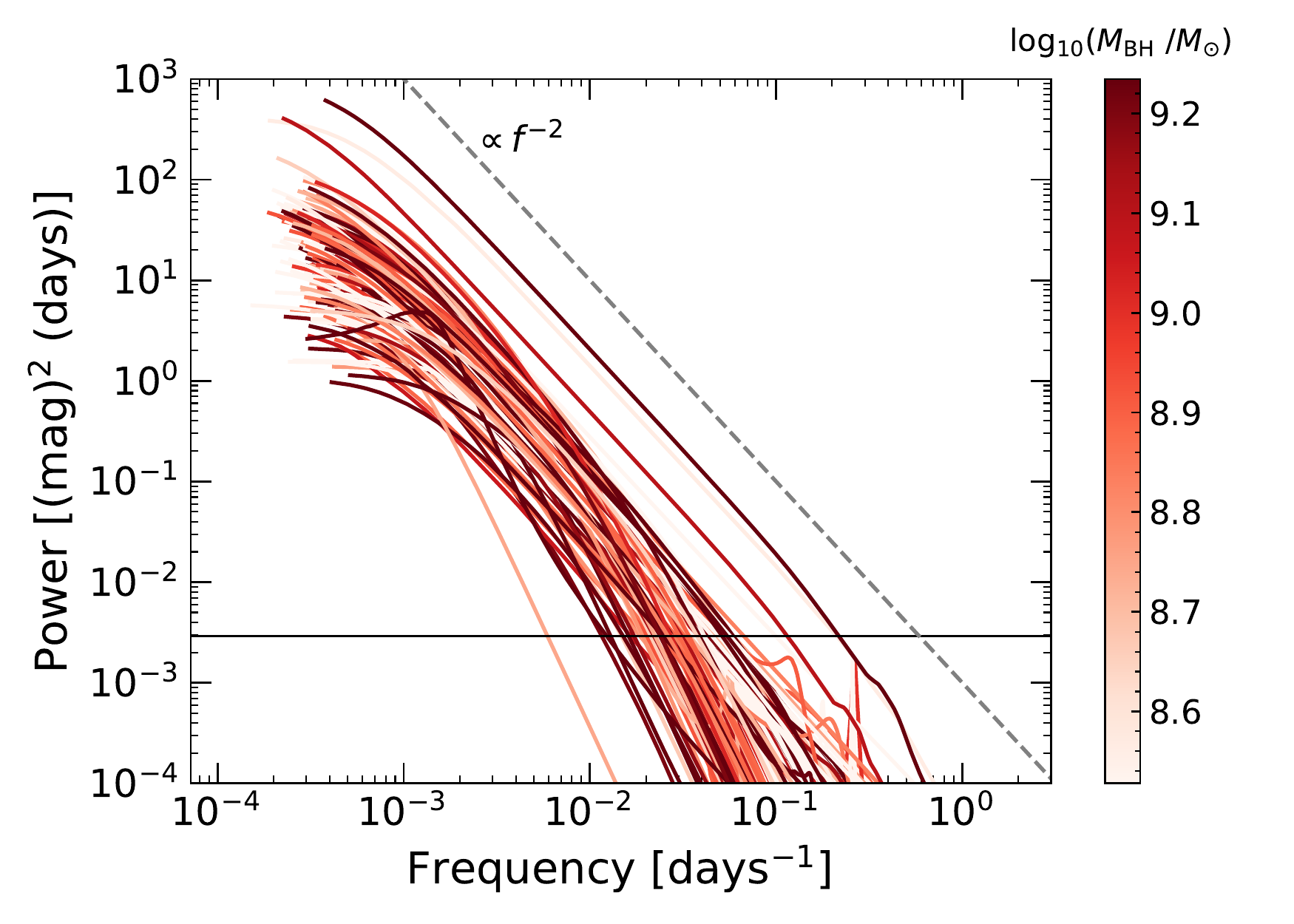}
    \caption{Rest-frame CARMA model PSDs ($g$-band) of all quasars in our parent sample, color-coded by different attributes of the target: $\log_{10}(\tau_{\rm DRW,rest})$, $\log_{10}(L_{\rm bol})$, and $\log_{10}(M_{\rm BH})$. In each panel, the black horizontal line represents the median noise level of the individual light curves, and the dashed gray line represents a $\propto f^{-2}$ PSD. There are some general trends visible, e.g., lower PSD amplitudes for higher-luminosity quasars, and faster flattening of the PSD at the low-frequency end for quasars with shorter damping timescales in the DRW fits. }
    \label{fig:all_psds}
\end{figure}

Fig.~\ref{fig:all_psds} displays all CARMA PSDs for our sample in the rest-frame of the quasar (only showing the best-fit model), color-coded by different properties. While these individual PSDs overlap considerably given their measurement uncertainties, there are trends of the PSD amplitude and shape with luminosity and black hole mass of the quasar. In addition, the CARMA PSD tends to flatten out sooner at the low-frequency end for light curves with shorter $\tau_{\rm DRW}$, suggesting that the DRW fits are reasonable in constraining the long-term damping timescale. 

\begin{figure}
    \centering
    \includegraphics[width=.46\textwidth]{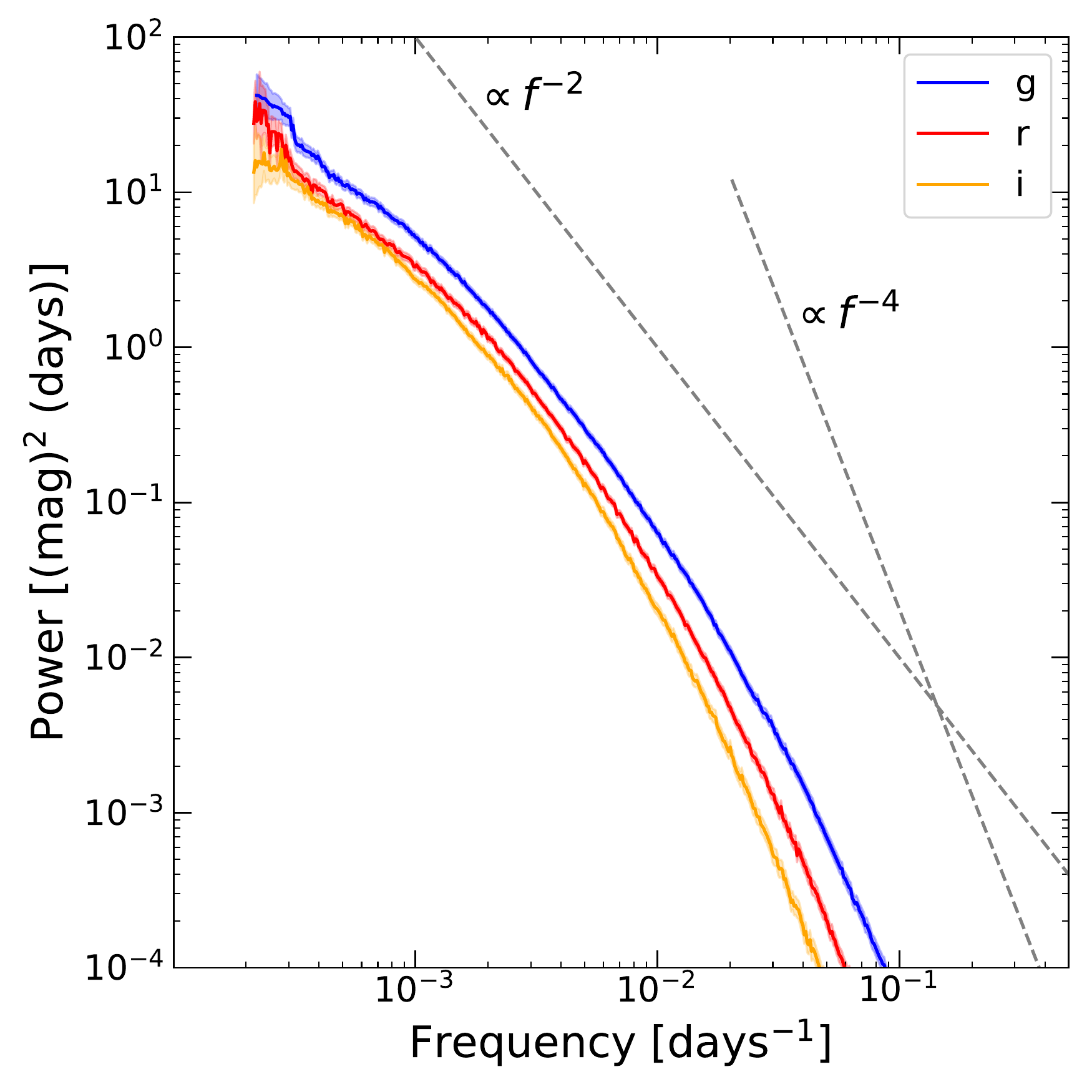}
    \caption{Rest-frame ensemble PSDs for the full sample in $gri$ bands. Each quasar light curve was fit using {\tt CARMA\_pack}, a code designed to fit time-series data to CARMA models using the method described in \citet{Kelly_etal_2014}, with optimized $(p,q)$ parameters for the CARMA model. For each PSD, the darker line shows the median value from the ensemble, and the light shaded area (nearly negligible at $f>10^{-3}\,{\rm days}^{-1}$) indicates the nominal uncertainty of the ensemble PSD.}
    \label{fig:EnsPSD}
\end{figure}

\begin{figure*}
    \centering
    \includegraphics[width=\textwidth]{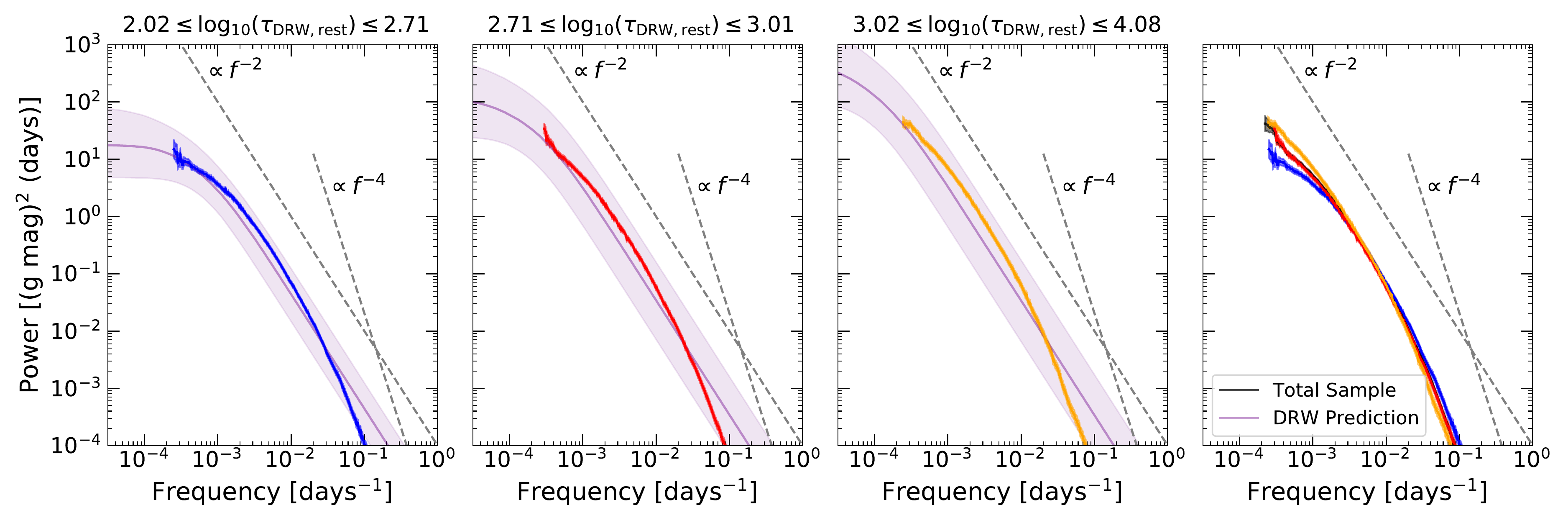}
    \caption{Ensemble CARMA PSDs for subsamples divided by their best-fit $\tau_{\rm DRW}$ from \S\ref{sec:drw}. The first three panels show these ensemble PSDs corresponding to each subsample, whose $\tau_{\rm DRW}$ ranges are shown above each panel. The DRW-predicted ensemble PSDs are shown in the purple-shaded area for each ensemble. The ensemble DRW-predicted PSDs are constructed in the same manner as the ensemble DRW-predicted structure functions, shown in Fig.~\ref{fig:3tauEnsSF}. The rightmost panel shows the PSDs of all three ensembles superimposed on the ensemble PSD for the full sample (shown in black). The two gray dashed lines in each panel correspond to $f^{-2}$ and $f^{-4}$ power laws. }
    \label{fig:3tauEnsPSD}
\end{figure*}

\begin{figure*}
    \centering
    \includegraphics[width=.9\textwidth]{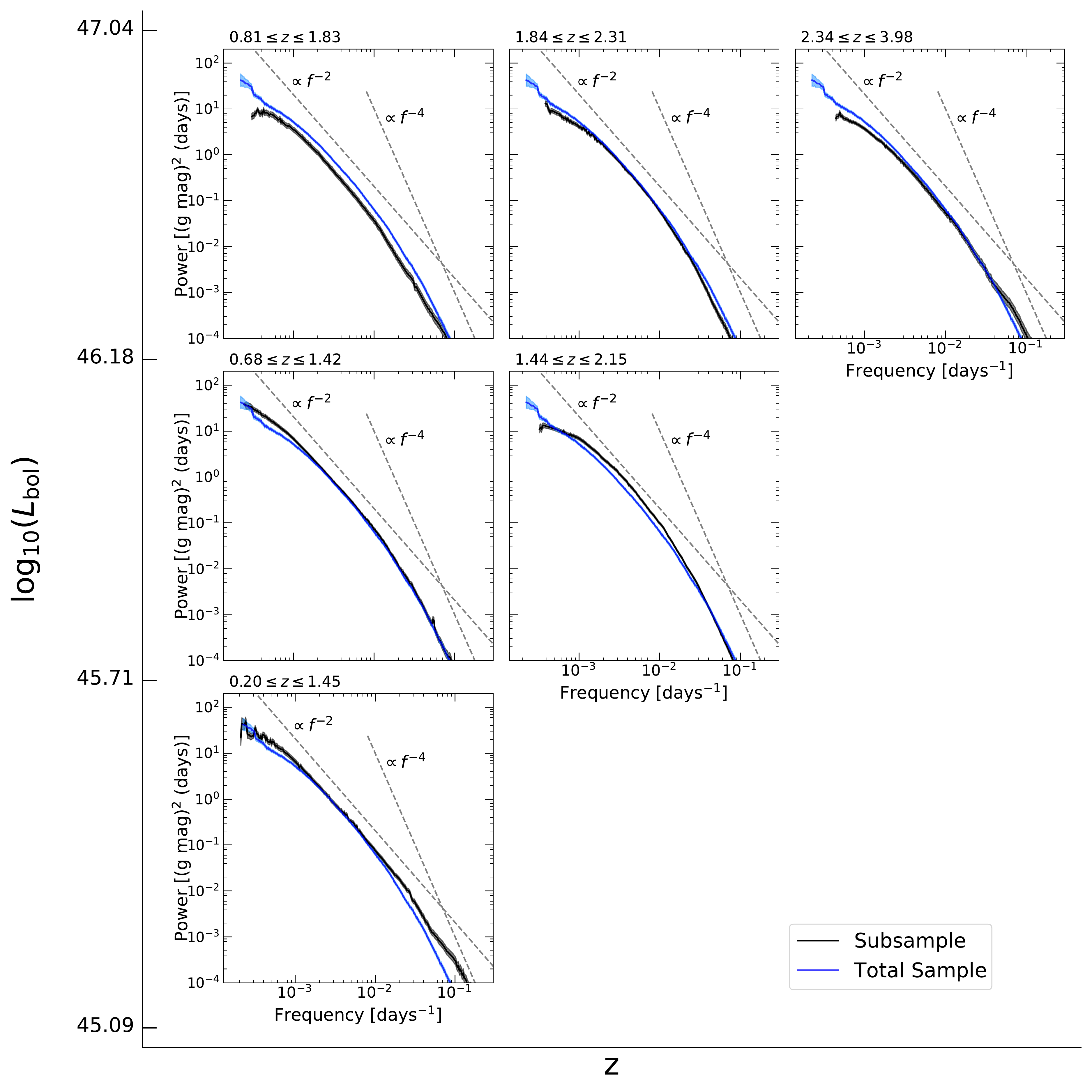}
    \caption{Similar to Fig.~\ref{fig:3tauEnsSF}, we group the full sample by $L_{\rm bol}$ and redshift and create ensemble PSDs. For each panel, the ensemble PSD is shown in blue, the ensemble PSD for the whole sample is shown in black, and the two gray dashed lines indicate a $f^{-2}$ PSD and a $f^{-4}$ PSD. The range of redshifts of each subsample is labeled above each panel, and the range of bolometric luminosities is indicated on the axis on the left. Subsamples in the same row have the same luminosity range (this is not true for the same column with slightly different redshift ranges). The artificial turnover of power at the lowest frequencies is caused by the limited number of objects ($\sim 10$) with the proper temporal coverage. }
    \label{fig:PSD_grid_g}
\end{figure*}

Fig.~\ref{fig:EnsPSD} shows the ensemble CARMA PSD for the full sample in the three bands. The ensemble PSDs are tightly constrained over days to decade timescales, and show a clear wavelength dependence. Fig.~\ref{fig:3tauEnsPSD} and Fig.~\ref{fig:PSD_grid_g} display the ensemble PSDs for the same subsets of quasars used in our SF analysis. When divided by the best-fit $\tau_{\rm DRW}$, the ensemble PSD agrees with the average DRW model in the subsample reasonably well, suggesting the DRW model provides a reasonable description of the underlying PSD. However, the more flexible CARMA model reveals a sharper decline in the variability power below timescales of a few weeks than the $f^{-2}$ power-law, consistent with earlier findings with other light curve samples \citep[e.g.,][]{Mushotzky_etal_2011,Zu_etal_2013}. In Appendix \ref{app:mock_lc}, we demonstrate that this PSD steepening at the highest frequencies is not due to the usage of a more flexible CARMA model or selection effects of our data, using simulated light curves. Similar to the SF analysis, the ensemble PSD shows dependences with wavelength and luminosity of the quasar, as shown in Fig.~\ref{fig:EnsPSD} and Fig.~\ref{fig:PSD_grid_g}.  


\begin{figure*}
    \centering
    \includegraphics[width=0.9\textwidth]{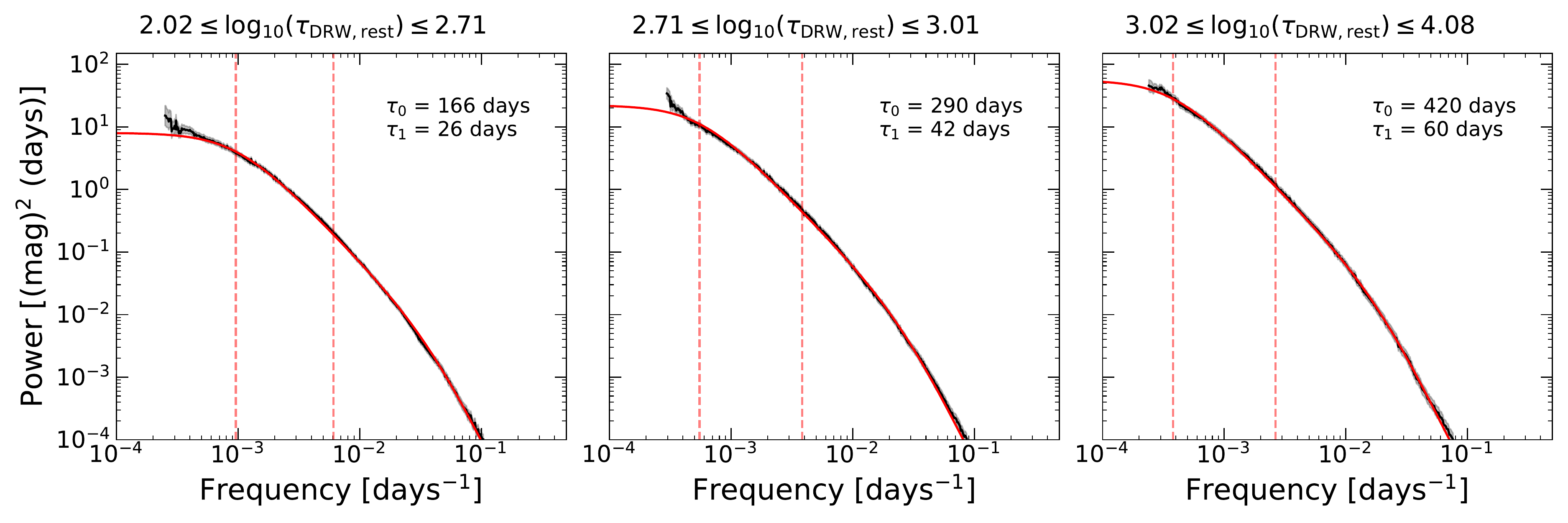}
    \caption{The ensemble PSD in the three subsamples divided by the best-fit $\tau_{\rm DRW,rest}$ (the black line), and a doubly-broken power-law fit (red line) with two break frequencies $f_0$ and $f_1$. The small measurement uncertainties in the ensemble PSD are shown in the gray shaded region (nearly invisible), representing the standard deviation in the PSD for each quasar, divided by the square root of the number of PSD data points in each frequency bin (i.e. $\sigma / \sqrt{N}$). The corresponding break timescales $\tau=1/(2\pi f)$ are marked in each panel. Both break timescales vary in concordance. }
    \label{fig:dbpl_fit}
\end{figure*}

\begin{figure}
    \centering
    \includegraphics[width=0.48\textwidth]{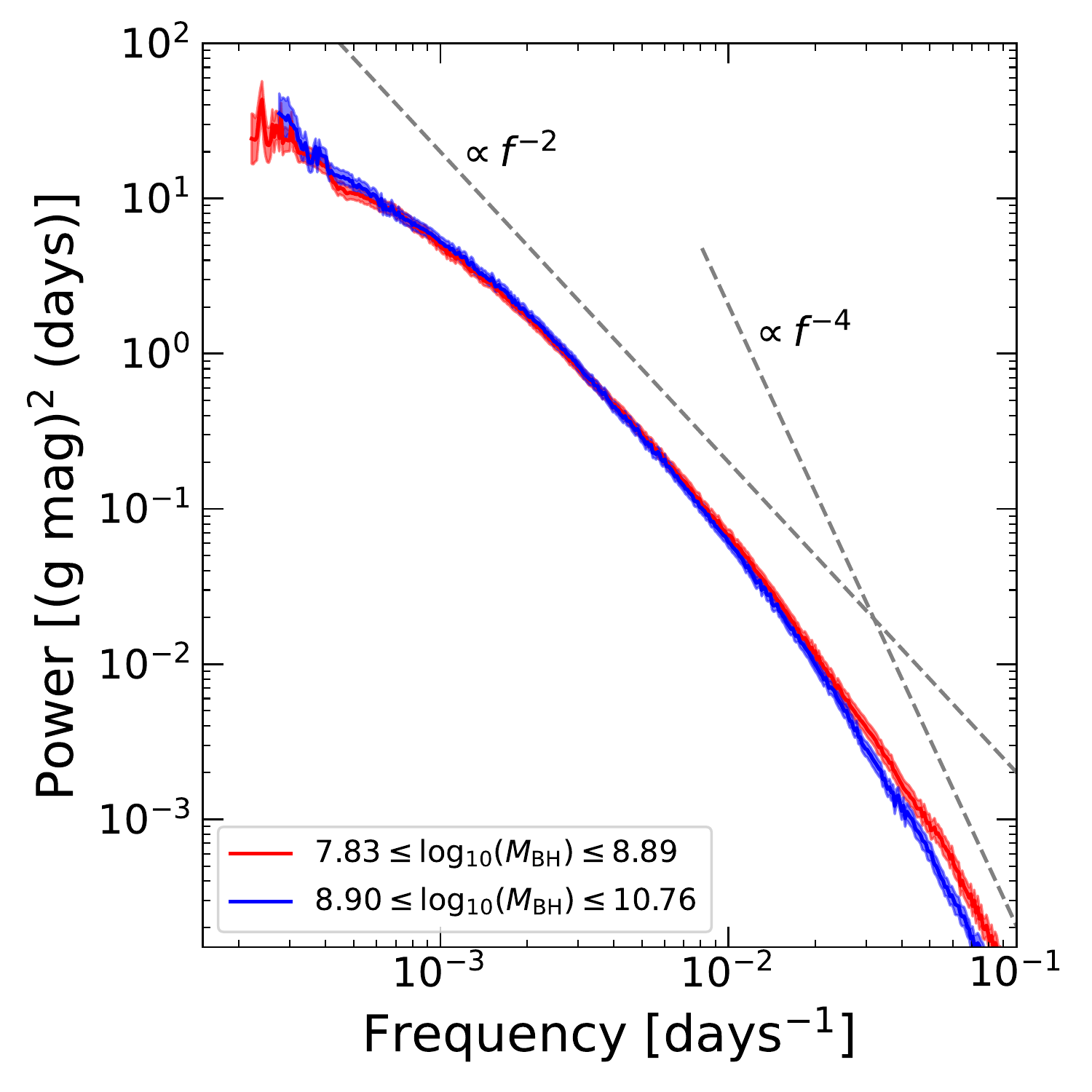}
    \caption{Ensemble PSDs for subsamples divided by the median virial black hole mass in the parent sample. While the dynamic range in black hole mass is limited in our sample, there is some evidence that the short-term break timescale is longer for the high-mass subsample. The mass dependence for the long-term (DRW) break timescale is less obvious, which would require a larger dynamic range in black hole mass \citep[e.g.,][]{Burke_etal_2021}.}
    \label{fig:LbolMbhEnsemblePSD}
\end{figure}

\section{Discussion}\label{sec:Disc}

\subsection{The wavelength dependence of $\tau_{\rm DRW}$}

We find that the DRW damping timescale only weakly depends on wavelength, consistent with earlier studies with shorter light curves. This weak wavelength dependence of the damping timescale is difficult to interpret: if $\tau_{\rm DRW}$ tracks the local timescale of the accretion disk, e.g., the thermal timescale, then we expect a stronger wavelength dependence of this timescale because the local thermal timescale scales with the emitting wavelength as $\tau\propto \lambda^{2}$ in the standard $\alpha$-disk model \citep{Shakura_Sunyaev_1973}. One possibility, as suggested by \citet{Burke_etal_2021}, is that the observed UV/optical variability is driven by processes in the inner (UV-emitting) part of the accretion disk, which rapidly propagates outwards at the Alfv\'{e}n speed, during which the characteristic variability timescale is more or less preserved. Alternatively, the observed damping timescale may be the thermal timescale averaged over different radii, leading to a shallower wavelength dependence \citep[e.g.,][]{Sun_etal_2020}. Further development of these theoretical ideas, combined with dedicated global radiation MHD simulations \citep[e.g.,][]{Jiang_etal_2019} will shed light on the nature of this long-term characteristic variability damping timescale. 

\subsection{Validity of the DRW prescription}

Overall, we find that the DRW model, even though an empirical prescription, describes the stochastic optical quasar light curves reasonably well over rest-frame timescales from a few months to a few years. The qualitative agreement between the DRW model and SF/PSD measurements suggests that the long-term characteristic variability timescale captured by the DRW model is reliable on average. Indeed, \citet{Burke_etal_2021} tested DRW fits to non-DRW light curves with a characteristic long-term turnover timescale in the PSD and found that the best-fit $\tau_{\rm DRW}$ correctly recovers this timescale. 

However, the length of the light curve will affect the constraints on $\tau_{\rm DRW}$ in a DRW fit \citep{Kozlowski_2017}. With our 20-yr baseline, we find that the median $\tau_{\rm DRW,rest}$ for S82 quasars is $\sim 750\,$days in the $g$ band, longer than the median $g$-band $\tau_{\rm DRW,rest}$ of 450 and 470 days if we use the shorter SDSS-only or SDSS+PS1 light curves. For comparison, \citet{MacLeod_etal_2010} and \citet{Suberlak_etal_2021} report a median $r$-band $\tau_{\rm DRW, rest}$ of 570 days using SDSS-only light curves. For $r$-band and using the SDSS-only light curves, we measure a median $\tau_{\rm DRW, rest}$ of $\sim 540$ days, consistent with these earlier studies. Details in the adopted ``best-fit'' DRW parameters and the MCMC convergence criterion in the fitting do not seem to impact our results much (see Appendix \ref{app:Sampling}). It is unclear if this is because some quasars have much longer intrinsic $\tau_{\rm DRW}$ than what can be realistically constrained by our current light curves, or because the light curve cannot be described by a single DRW process. For example, if the quasar light curve contains a long-term gradual trend in addition to a DRW process, the best-fit $\tau_{\rm DRW}$ will increase as the baseline increases (Appendix \ref{app:DRW}). It is possible that the accretion state of the quasar is gradually changing over multi-year timescales \citep[e.g.,][]{Caplar_etal_2020}, leading to long-term trends in the light curve and biasing the DRW fit that assumes stationarity. Continued monitoring of these quasars in our dedicated DECam program will address this question with even more extended light curves. 


On timescales shorter than $\sim$ a month, however, the slope of the PSD is noticeably steeper than $-2$. In Appendix \ref{app:mock_lc}, we use simulated DRW light curves matched to the observed cadences and S/N of our sample to test if a more flexible CARMA fit would lead to an artificially steeper high-frequency-end slope. We find that the resulting CARMA PSD has a high-frequency-end slope of $-2$, which confirms that the steeper PSD slope observed in our sample is real. While the exact asymptotic slope of the PSD is likely impacted by the CARMA model restrictions, the locations of the slope transitions are largely determined by the data. There is evidence (e.g., Fig.~\ref{fig:3tauEnsPSD}) that this short timescale cutoff of power is positively correlated with the long-term damping timescale. To further illustrate this point, we fit a doubly-broken power-law model to the three ensemble PSDs divided by the measured $\tau_{\rm DRW,rest}$ in Fig.~\ref{fig:3tauEnsPSD}: $P\propto 1/[1 + (f/f_0)^2 + (f/f_1)^4]$. This PSD model fits the three ensemble PSDs reasonably well over years to days timescales, as shown in Fig.~\ref{fig:dbpl_fit}. The two break timescales, $\tau_0=1/(2\pi f_0)$ and $\tau_1=1/(2\pi f_1)$, indeed vary in concordance in the three ensembles.

While our sample is small and the dynamic range in black hole mass or quasar luminosity is limited, there is also some tentative evidence that this high-frequency-end break occurs at shorter timescales for lower-luminosity (and less massive) quasars (Fig.~\ref{fig:PSD_grid_g}). This point is further illustrated in Fig.~\ref{fig:LbolMbhEnsemblePSD}, where we compare the ensemble PSDs for subsamples divided by black hole mass. If we assume both break timescales scale with black hole mass as $M_{\rm BH}^{0.5}$ \citep{Burke_etal_2021}, we expect much shorter high-frequency break timescales in low-redshift Seyferts ($M_{\rm BH}\sim 10^7\,M_\odot$) than in SDSS quasars ($M_{\rm BH}\sim 10^9\,M_\odot$). This may explain the much shorter (a few days) cutoff timescales found for low-redshift, low-luminosity AGNs that are two orders of magnitude less massive than SDSS quasars \citep[e.g.,][]{Mushotzky_etal_2011}. 

The physical origin of the suppression of variability power on timescales shorter than $\sim 1$ month is unclear. It could be due to the intrinsic shape of the variability PSD, e.g., resulting from the break in the driving variability PSD and/or damping processes in the accretion disk \citep[e.g.,][]{Sun_etal_2020}. An alternative explanation, as pointed out by, e.g., \citet{Tachibana_etal_2020}, is due to an averaging effect. Even if the flux of the accretion disk varies coherently, emission from different parts of the disk or from more spatially-extended regions (e.g., an extended diffuse continuum emission region or the broad-line region) will arrive at different times. The observed variable flux is then the convolution of the intrinsic variability pattern with the transfer function describing the time delays from different locations. \citet{Tachibana_etal_2020} showed that, with a likely transfer function form (a semi-circle with a characteristic timescale of a month), the short-time variability power will be reduced due to averaging, producing a PSD slope close to $-4$ beyond this characteristic frequency. In general, such transfer functions would reduce the high-frequency power, leading to a steeper high-frequency end slope in the observed PSD. In both the intrinsic PSD scenario and the ``smearing'' scenario, it is possible that the characteristic timescale of this second high-frequency-end break, which reflects some characteristic size of the emission region, depends on the physical properties of the quasar, such as the black hole mass \citep{Sun_etal_2020,Tachibana_etal_2020}, in a similar way as the long-term damping timescale $\tau_{\rm DRW}$.



\section{Conclusions} \label{sec:Conc}

Given the simplicity of the DRW model and its reasonable success to fit quasar light curves, it has become increasingly popular to use the DRW prescription to describe stochastic quasar variability. However, the validity of the DRW prescription has to be tested with high-quality light curve data that are well sampled, have sufficient baselines and adequate S/N. Some recent light curve samples already have sufficient quality to reveal evidence for deviations from the DRW prescription either for individual objects or for large quasar samples \citep[e.g.,][]{Mushotzky_etal_2011,Zu_etal_2013,Kasliwal_etal_2015,Caplar_etal_2017,Yu_etal_2022}.

In this work we have measured the optical continuum variability of a sample of 190 quasars from the SDSS Stripe 82 region. Our quasar sample has been photometrically monitored in the SDSS, PS1, DES surveys, as well as our continued monitoring with DECam. The light curves of our sample span a baseline of $\sim 20$ years with $\sim 200$ epochs in each of the $gri$ bands. We fit these light curves with the DRW model, and measured the structure function and power spectrum density using the CARMA models. The main findings from our work are the following:

\begin{enumerate}
    \item[$\bullet$] The best-fit DRW parameters ($\tau_{\rm DRW}$ and ${\rm SF}_{\infty}$) continue to rise with our light curve data, compared with earlier studies with shorter (e.g., 10-yr and 15-yr) baselines from SDSS-only \citep{MacLeod_etal_2010} and SDSS+PS1 \citep{Suberlak_etal_2021}. The average rest-frame $\tau_{\rm DRW}\sim 750$\,days in $g$ band for S82 quasars with our 20-yr light curves. 
    
    
    \item[$\bullet$] While the $\tau_{\rm DRW}$ measurements for many S82 quasars are still not well constrained with the 20-yr light curves, we believe that the bias from insufficient baselines is reduced compared with earlier studies based on shorter baselines, if the underlying variability process is indeed a single DRW. However, we caution that realistic quasar light curves may be more complicated than a single DRW process, e.g., multiple variability processes with different characteristic timescales and/or non-stationary variability processes could be at work. In such cases, the results from a single DRW fit will depend on the baseline. More extended light curves are required to test this possibility.   
    
    \item[$\bullet$] Using a subset of 27 quasars for which we have relatively better-constrained $\tau_{\rm DRW}$ in the $g$, $r$ and $i$ bands, we confirm a weak wavelength dependence of $\tau_{\rm DRW}\propto \lambda^{0.51\pm0.20}$ ($\tau_{\rm DRW}\propto \lambda^{0.34\pm0.10}$ for the full sample). This wavelength dependence is slightly stronger than previous results $\tau_{\rm DRW}\propto \lambda^{0.17}$ based on 10-yr light curves \citep{MacLeod_etal_2010}, although these results are formally consistent within 2$\sigma$. 
    
    \item[$\bullet$] We also measured the optical SF and PSD of our quasar sample. The baseline and sampling of our light curves enabled reliable constraints of the ensemble PSD over days to decades timescales. Comparisons between the ensemble SF and PSD with predictions from the best-fit DRW models suggest that the DRW prescription provides a reasonably good description of the variability properties of quasars over months to years timescales. But the average PSD slope on timescales shorter than $\sim$ a month is noticeably steeper than the DRW model, consistent with earlier findings \citep[e.g.,][]{Mushotzky_etal_2011,Zu_etal_2013}. There is tentative evidence that this high-frequency cutoff timescale correlates with the low-frequency damping timescale $\tau_{\rm DRW}$, hence both timescales may have similar dependences on physical properties of the quasar \citep[e.g.,][]{Burke_etal_2021}. 
    
\end{enumerate}

We continue to monitor our quasar sample during 2020-2024 as part of our ongoing effort to photometrically monitor deep extragalactic fields with ample multi-wavelength and time-domain data. With another $\sim 5$ year extension of the baseline and seamlessly merging with light curves from the Vera C. Rubin Observatory Legacy Survey of Space and Time \citep{Ivezic_etal_2019}, this quasar sample will become a prime sample to study quasar optical continuum variability. Such studies will further test the applicability of the DRW model and the stationarity of the stochastic variability process, as well as provide insights on the physical origin of quasar variability.

\section*{Acknowledgements}

We thank the anonymous referee for useful comments that improved the manuscript, Mouyuan Sun for useful discussions on the CHAR model, and Brad Tucker and Manda Banerji for useful comments on the draft. ZS and YS acknowledge support from NSF grants AST-1715579 and AST-2009947, and NASA grant 80NSSC21K0775. 

This research made use of matplotlib, a Python library for publication quality graphics \citep{matplotlib_2007}. This research made use of SciPy \citep{scipy_2020}. This research made use of Astropy, a community-developed core Python package for Astronomy \citep{astropy_2018}. This research made use of NumPy \citep{numpy_2020}. This research made use of pandas \citep{pandas_2011}. 

Some of the data presented in this paper were obtained from the Mikulski Archive for Space Telescopes (MAST). STScI is operated by the Association of Universities for Research in Astronomy, Inc., under NASA contract NAS5-26555. Support for MAST for non-HST data is provided by the NASA Office of Space Science via grant NNX13AC07G and by other grants and contracts. 


Funding for the SDSS and SDSS-II has been provided by the Alfred P. Sloan Foundation, the Participating Institutions, the National Science Foundation, the U.S. Department of Energy, the National Aeronautics and Space Administration, the Japanese Monbukagakusho, the Max Planck Society, and the Higher Education Funding Council for England. The SDSS Web site is http://www.sdss.org/.

The SDSS is managed by the Astrophysical Research Consortium for the Participating Institutions. The Participating Institutions are the American Museum of Natural History, Astrophysical Institute Potsdam, University of Basel, University of Cambridge, Case Western Reserve University, University of Chicago, Drexel University, Fermilab, the Institute for Advanced Study, the Japan Participation Group, Johns Hopkins University, the Joint Institute for Nuclear Astrophysics, the Kavli Institute for Particle Astrophysics and Cosmology, the Korean Scientist Group, the Chinese Academy of Sciences (LAMOST), Los Alamos National Laboratory, the Max-Planck-Institute for Astronomy (MPIA), the Max-Planck-Institute for Astrophysics (MPA), New Mexico State University, Ohio State University, University of Pittsburgh, University of Portsmouth, Princeton University, the United States Naval Observatory, and the University of Washington.

Funding for the DES Projects has been provided by the U.S. Department of Energy, the U.S. National Science Foundation, the Ministry of Science and Education of Spain, the Science and Technology Facilities Council of the United Kingdom, the Higher Education Funding Council for England, the National Center for Supercomputing Applications at the University of Illinois at Urbana-Champaign, the Kavli Institute of Cosmological Physics at the University of Chicago, 
the Center for Cosmology and Astro-Particle Physics at the Ohio State University,
the Mitchell Institute for Fundamental Physics and Astronomy at Texas A\&M University, Financiadora de Estudos e Projetos, Funda{\c c}{\~a}o Carlos Chagas Filho de Amparo {\`a} Pesquisa do Estado do Rio de Janeiro, Conselho Nacional de Desenvolvimento Cient{\'i}fico e Tecnol{\'o}gico and the Minist{\'e}rio da Ci{\^e}ncia, Tecnologia e Inova{\c c}{\~a}o, the Deutsche Forschungsgemeinschaft and the Collaborating Institutions in the Dark Energy Survey. 

The Collaborating Institutions are Argonne National Laboratory, the University of California at Santa Cruz, the University of Cambridge, Centro de Investigaciones Energ{\'e}ticas, Medioambientales y Tecnol{\'o}gicas-Madrid, the University of Chicago, University College London, the DES-Brazil Consortium, the University of Edinburgh, the Eidgen{\"o}ssische Technische Hochschule (ETH) Z{\"u}rich, 
Fermi National Accelerator Laboratory, the University of Illinois at Urbana-Champaign, the Institut de Ci{\`e}ncies de l'Espai (IEEC/CSIC), 
the Institut de F{\'i}sica d'Altes Energies, Lawrence Berkeley National Laboratory, the Ludwig-Maximilians Universit{\"a}t M{\"u}nchen and the associated Excellence Cluster Universe, the University of Michigan, NSF's NOIRLab, the University of Nottingham, The Ohio State University, the University of Pennsylvania, the University of Portsmouth, SLAC National Accelerator Laboratory, Stanford University, the University of Sussex, Texas A\&M University, and the OzDES Membership Consortium.

Based in part on observations at Cerro Tololo Inter-American Observatory at NSF's NOIRLab (NOIRLab Prop. ID 2012B-0001; PI: J. Frieman), which is managed by the Association of Universities for Research in Astronomy (AURA) under a cooperative agreement with the National Science Foundation.

The DES data management system is supported by the National Science Foundation under Grant Numbers AST-1138766 and AST-1536171.
The DES participants from Spanish institutions are partially supported by MICINN under grants ESP2017-89838, PGC2018-094773, PGC2018-102021, SEV-2016-0588, SEV-2016-0597, and MDM-2015-0509, some of which include ERDF funds from the European Union. IFAE is partially funded by the CERCA program of the Generalitat de Catalunya. Research leading to these results has received funding from the European Research Council under the European Union's Seventh Framework Program (FP7/2007-2013) including ERC grant agreements 240672, 291329, and 306478. We acknowledge support from the Brazilian Instituto Nacional de Ci\^encia
e Tecnologia (INCT) do e-Universo (CNPq grant 465376/2014-2).

This manuscript has been authored by Fermi Research Alliance, LLC under Contract No. DE-AC02-07CH11359 with the U.S. Department of Energy, Office of Science, Office of High Energy Physics.

Based in part on observations at Cerro Tololo Inter-American Observatory at NSF's NOIRLab (NOIRLab Prop. ID 2019B-0219; PI: X. Liu), which is managed by the Association of Universities for Research in Astronomy (AURA) under a cooperative agreement with the National Science Foundation.

The Pan-STARRS1 Surveys (PS1) have been made possible through contributions of the Institute for Astronomy, the University of Hawaii, the Pan-STARRS Project Office, the Max-Planck Society and its participating institutes, the Max Planck Institute for Astronomy, Heidelberg and the Max Planck Institute for Extraterrestrial Physics, Garching, The Johns Hopkins University, Durham University, the University of Edinburgh, Queen's University Belfast, the Harvard-Smithsonian Center for Astrophysics, the Las Cumbres Observatory Global Telescope Network Incorporated, the National Central University of Taiwan, the Space Telescope Science Institute, the National Aeronautics and Space Administration under Grant No. NNX08AR22G issued through the Planetary Science Division of the NASA Science Mission Directorate, the National Science Foundation under Grant No. AST-1238877, the University of Maryland, and Eotvos Lorand University (ELTE). 

\section*{Data Availability}

We provide all light curve data and time series measurements in two online FITS tables located at \url{https://zenodo.org/record/5842449#.YipOg-jMJPY}. The format of these FITS tables is described in Tables \ref{tab:catalogall} and \ref{tab:catalog_sf_psd}.




\bibliographystyle{mnras}
\bibliography{Refs1,Refs2} 




\begin{appendix}

\section{Light Curve Analysis}\label{sec:appendix}

\subsection{Structure Function}\label{app:sf}

One of the more traditional ways of modeling the variability of quasars is through structure function analysis. This method describes the change in magnitude as a function of time lag $\Delta t$ between two observations. Since the SF calculation is model-independent, it provides an empirical view of quasar variability with no underlying assumptions. The most basic way to define the structure function is the root-mean-square magnitude difference for a given grid of time lags.



However, without accounting for the flux measurement uncertainties, structure function measurements at small $\Delta t$ will level off to a certain ``SF floor''. Therefore, using the method described by \citet{Kozlowski_2017}, we subtract the measurement errors of both observations in the pair in quadrature:

\begin{equation}
    SF(\Delta t) = \sqrt{ \frac{1}{N_{\Delta t}} \sum_{i<j}\left(  (y_i - y_j)^2 - \sigma_i^2 - \sigma_j^2 \right) }\ ,
\end{equation}\label{eqn:MeasureSF}
where $\sigma_i$ and $\sigma_j$ are the measurement errors in observations $y_i$ and $y_j$ respectively.

The structure function is related to the auto-correlation function (ACF) of the light curve. Assuming that the variability of the source is stationary, we can use the covariance of two signals to compute the structure function:

\begin{equation}
    SF(\Delta t) = \sqrt{ 2\sigma_s^2 (1 - ACF(\Delta t)) }\ ,
\end{equation}
where $\sigma_s$ is the variability amplitude intrinsic to the source. Taking the limit as $\Delta t \rightarrow \infty$, we obtain:

\begin{equation}
    SF(\Delta t) = SF_{\infty} \sqrt{ 1 - ACF(\Delta t) }\ ,
\end{equation}
where $SF_{\infty}^2\equiv 2\sigma_s^2$ is the value of the structure function as $\Delta t \rightarrow \infty$. Assuming the variability is stationary (meaning the mean value of the light curve does not change), the difference between signals at large time lags will approach a constant value proportional to the intrinsic variability amplitude (white noise). The structure function will also flatten to white noise at very short time lags, where the change in magnitude is on the order of the measurement uncertainty.


We utilize Eqn.~\ref{eqn:MeasureSF} to make all of our structure function measurements, where time lags are shifted to the rest-frame of the quasar. We also make ensemble structure function measurements for various subsamples of our 190 quasar dataset. To derive ensemble structure functions from individual objects, we bin the structure functions of each individual object into the same $\Delta t$ grid. We then take the median of each bin to be the ensemble measurement for that time lag, and use the uncertainty on the median (the standard deviation of the samples in each bin, divided by $\sqrt{N}$ the number of samples in the bin) to represent the uncertainty in that measurement. We create these ensemble structure functions for the total sample, three subsamples grouped by their fitted $\tau_{\rm DRW}$, and six subsamples grouped by their bolometric luminosity and redshift.

For \emph{g}-band measurements, through visual inspection, we observed that the structure function began to rise near time lags of 10 days. However, when measuring these ensemble structure functions, we noticed that they started to flatten at time lags less than days to weeks in the quasar rest frame). This proved to be more prevalent for structure functions in the \emph{r} and \emph{i} bands, where the structure function would be almost constant until $\Delta t_{\rm rest} \sim$5-10 days and then jump. We attribute this flattening to PSF seeing variations on short timescales, measurement uncertainties, as well as host galaxy contamination. To minimize the effect of this flattening, we perform linear regression (in log-space) on this floor using the method of \citet{Kelly_2007} and subtract the best-fit line from the full ensemble structure function. This floor stopped at different time lags for each band, [5,20,40] days for \emph{gri} measurements respectively, which we use to set the linear regression range.

For each ensemble structure function, excluding for the total sample, we also overlay the DRW model prediction for comparison. The DRW-predicted structure functions are also ensembles, using the predicted $\tau_{\rm DRW}$ and $\sigma_{\rm DRW}$ and Eqn.~\ref{eqn:PSDDRW}. These ensemble DRW-predicted structure functions are obtained in a similar manner to the ensemble structure functions themselves: we create structure functions for each target in the ensemble using the best-fit DRW parameters, then we bin the structure functions onto a common $\Delta t$ grid. We then take the median of each bin to get an ensemble DRW-predicted structure function.

\subsection{The Damped Random Walk Model}\label{app:DRW}

The Damped Random Walk model, also known as the Ornstein-Uhlenbleck process, is a statistical model used to describe the stochastic variability from the accretion disk emission of quasars. This Gaussian process is the simplest model of a family of Gaussian processes known as continuous auto-regressive moving-average (CARMA) models. General CARMA models, discussed in \S \ref{app:psd}, specify that the output of the model is linear in the current and past terms in the time-series. This is seen in the DRW model (a CARMA(1,0) model), as it has a term that pushes large deviations from the mean of the time-series back towards the mean. It is useful to model light curves with the DRW model as it has parameters that can potentially connect to physical parameters of the quasar, and it can be modelled directly in the time domain instead of the frequency domain. Quasar variability studies in the frequency domain are subject to windowing effects, as large gaps in the data can lead to power leakage and aliases. Using a DRW model (or any CARMA model) can mediate these adverse effects.

All Gaussian processes require a covariance matrix (also known as a kernel), governing the relationship between two points in a time series. In the case of a DRW process, the covariance matrix is
\begin{equation}\label{eqn:kernel}
    k(t_{nm}) = \sigma_{\rm DRW}^2 \exp(-t_{nm} / \tau_{\rm DRW})\ ,
\end{equation}
where $t_{nm} = |t_n - t_m|$ and $t_n$, $t_m$ are times within the time series. $\sigma$ is the long-term standard deviation of variability, and $\tau$ defines a characteristic timescale where the PSD of this process breaks. We can relate this model to the structure function and the PSD in the following way:
\begin{equation}\label{eqn:SFDRW}
    SF^2(\Delta t) = 2\sigma_{\rm DRW}^2 \left( 1 - e^{-|\Delta t|/\tau_{\rm DRW}}  \right)\ ,
\end{equation}
\begin{equation}\label{eqn:PSDDRW}
    P(f) = \frac{4 \sigma_{\rm DRW}^2 \tau_{\rm DRW}}{1 + (2\pi f \tau_{\rm DRW})^2}\ ,
\end{equation}
where $P(f)$ is the PSD as a function of frequency $f$. By comparing Eqn.~\ref{eqn:SFDRW} and Eqn.~\ref{eqn:MeasureSF}, we have $SF_{\infty}^2 = 2\sigma_{\rm DRW}^2$ and $ACF(\Delta t) = \exp(-|\Delta t|/\tau_{\rm DRW})$. This PSD follows white noise at low frequencies ($\propto f^0$), and transitions to a $f^{-2}$ PSD at higher frequencies below the characteristic timescale $\tau_{\rm DRW}$.



In this study, we model our quasar light curves using the fast Gaussian process solver {\tt Celerite} \citep{Celerite}, which uses Gaussian process regression to fit the time series to a specified kernel. Given a number of terms in the kernel, and a method to maximize, {\tt Celerite} can fit a time series to derive the best-fit parameters to said kernel. In our case, we utilize a DRW kernel (specified in Eqn.~\ref{eqn:kernel}), as well as a term to characterize the effect of a white noise floor from unknown systematic flux errors ($\sigma_n$), also called jitter. We use uniform priors on all parameters within the input {\tt Celerite} kernel (in log-space), and allow {\tt Celerite} to minimize the log-likelihood in parameter-space to obtain a set of parameters to fit the light curve. We then use the MCMC sampler {\tt emcee} \citep{Emcee} implemented in Python to draw from the joint posterior probability distribution output from {\tt Celerite}. The final parameters compiled in the FITS table described in Table~\ref{tab:catalogall} are the median samples from these MCMC samples. The upper and lower errors for these parameters are obtained from the 16$^{\rm th}$ and 84$^{\rm th}$ percentiles of the samples. One example {\tt Celerite} fit is shown in Fig.~\ref{fig:taufit}.


\begin{figure}
    \centering
    \includegraphics[width=0.5\textwidth]{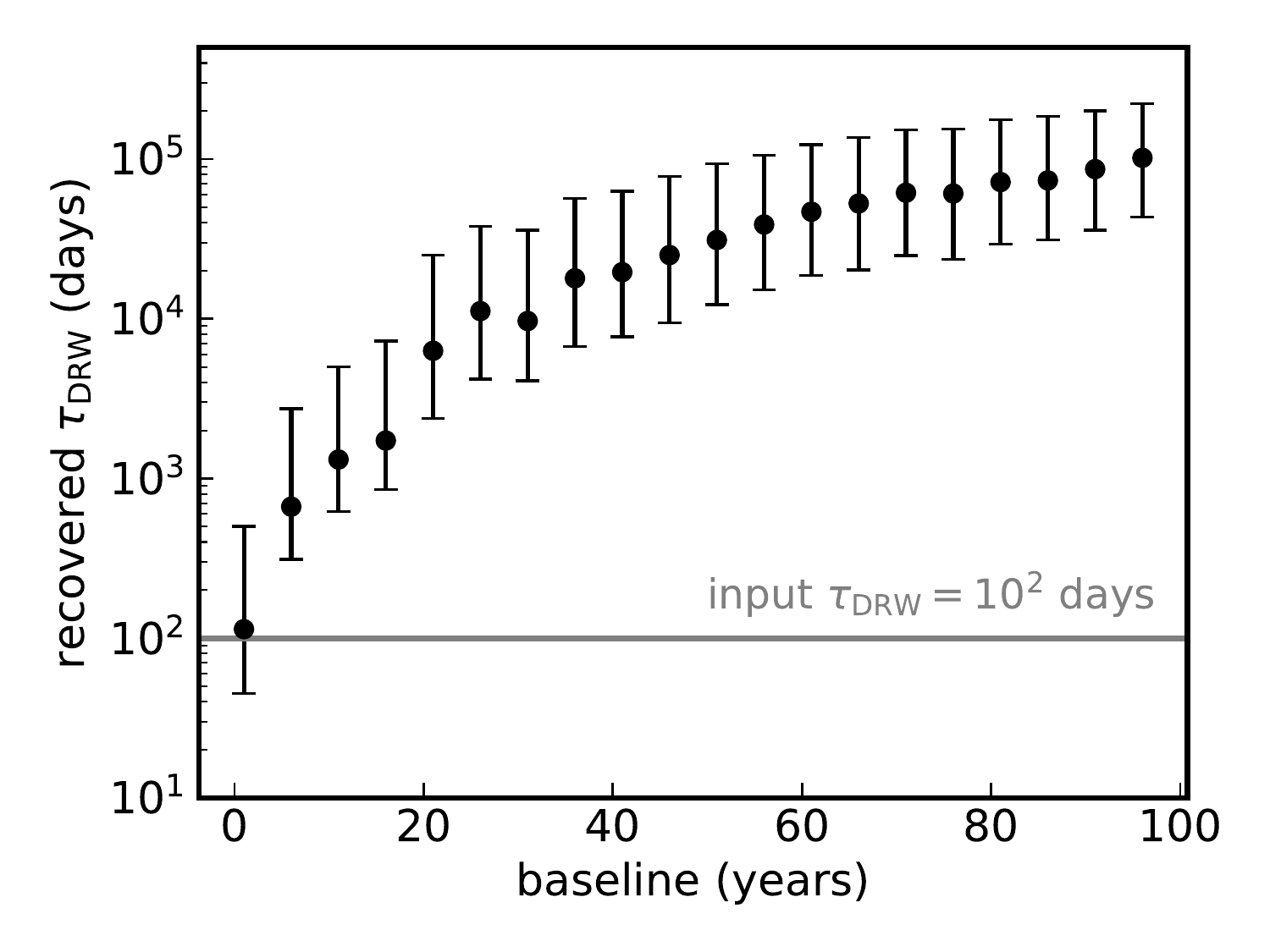}
    \caption{Recovered damping timescales for simulated non-stationary DRW light curves with input $\tau_{\rm{DRW}}=100$ days plus a linear trend of $0.0365$ mag yr$^{-1}$ at varying baselines.}
    \label{fig:drw_trending}
\end{figure}

However, there are potentially additional features in the light curve that can skew the results of the DRW fit. Here, we investigate the effects of a long-term trend in the light curve on the recovery of $\tau_{\rm DRW, obs}$ using {\tt Celerite} and simulated data. We input simulated DRW light curves with input $\tau_{\rm DRW,obs} = 100$ days, but add a long-term linear trend (\emph{non-stationarity}) to the light curve, in this case of $1 \times 10^{-4}$ mag per day. We generate mock light curves using this hybrid model with different baselines, and use {\tt Celerite} to extract a $\tau_{\rm DRW, obs}$ from the simulated light curve. The results (shown in Fig.~\ref{fig:drw_trending}), show that as the baseline of the non-stationary light curve increases, the extracted $\tau_{\rm DRW, obs}$ increases as well. In this test, the input $\tau_{\rm DRW,obs}$ is 100 days, and is reasonably recovered for short baselines (less than $\sim 10$ years). However, as the baseline increases, the linear long-term trend starts to skew the recovery of $\tau_{\rm DRW,obs}$ towards longer and longer damping timescales.


\subsection{PSD Analysis with CARMA models}\label{app:psd}

While many studies have shown that the DRW model can describe quasar light curve variability to a reasonable degree, we understand that it is not the only model available. It has been shown that stochasic processes generated from non-DRW models can be modeled with DRW \citep{Kozlowski_2016}, albeit with biased DRW parameters. Therefore, to get a true sense of the PSD of quasar light curves and the stochastic processes occurring within their accretion disks, we utilize the more general CARMA model to obtain PSD measurements. Whereas DRW-modeled PSDs are restricted to having a white noise at low frequencies and a $f^{-2}$ PSD at higher frequencies (with a characteristic break timescale in between them), CARMA-predicted PSDs are not restricted to such a shape.

The PSD of a CARMA model is described in the following manner:
\begin{equation}\label{eqn:carma_psd}
    P(f) = \sigma^2 \frac{| \sum_{j=0}^{q} \beta_j (2\pi i f)^j |^2}{| \sum_{k=0}^{p} \alpha_k (2\pi i f)^k |^2}\ ,
\end{equation}
where $\sigma^2$ is the variance of the modeled white noise process, $\alpha_j$ are the auto-regressive parameters of the model, and $\beta_k$ are the moving-average parameters of the model. The order of the CARMA model is defined by the $p$ and $q$ parameters, which define the number of auto-regressive and moving-average components respectively. The requirement that CARMA processes are stationary also requires that $q < p$. By convention, we set $\beta_0 =1$ and $\alpha_p = 1$. When setting $p=1$ and $q=0$, we recover the DRW PSD, as well as the covariance matrix, where $\tau_{\rm DRW} = 1/\alpha_0$ and $\sigma_{\rm DRW} = \sigma \sqrt{\tau_{\rm DRW}}/2$.

While we have used {\tt Celerite} to fit the quasar light curves to the DRW model, we now opt to utilize the widely used {\tt CARMA\_pack} code \citep{Kelly_etal_2014} to fit our light curves to a generalized CARMA(p,q) model. While the covariance matrix for the DRW model is somewhat simple, it becomes increasingly complex as the order of the CARMA model is increased, and therefore increasingly more complex to implement into {\tt Celerite}. The generality of the kernel terms available in {\tt Celerite} allows the implementation of a large variety in the kernels that can be used, but formulating the CARMA PSD in terms of {\tt Celerite}'s kernel terms is highly involved. {\tt CARMA\_pack} also includes the functionality of choosing an optimal $(p,q)$ of the model used to fit the time series.


We perform the CARMA modeling using time series in the rest-frame of each quasar. To model our light cures to a generalized CARMA model with {\tt CARMA\_pack}, we obtain the optimal $(p,q)$ of the model. {\tt CARMA\_pack} does this by finding the maximum likelihood estimate of the CARMA models produced from a user-input grid of $(p,q)$ values. We choose to search a parameter space where $1 < p \leq 7$ and all $q < p$. After using 100 different optimizers initialized to random values for the CARMA parameters for a given model, the maximum likelihood estimate is chosen as the best-fit parameters for that model. This is process is performed for a specified region in parameter space of $p$ and $q$, after which the code picks the $(p,q)$ combination which minimizes the corrected Akaike Information Criterion \citep[AICc,][]{Akaike_1973} provided by \cite{Hurvich_1989}. After choosing the optimal CARMA model for a given object, we use {\tt CARMA\_pack} to derive the maximum likelihood posterior distribution for all of the CARMA parameters. We then use {\tt CARMA\_pack}'s MCMC implementation to sample the CARMA parameters, given the order of the model. After testing the effect of the number of iterations of the MCMC on the convergence of fitted parameters (discussed in Appendix \ref{app:Sampling}), we found the results are well convergent for 60,000 iterations and 30,000 burn-in samples. After running the MCMC sampler, {\tt CARMA\_pack} will then output samples for all of the CARMA parameters using the posterior distribution of the object's fitted CARMA model. We can then use {\tt CARMA\_pack} to sample the PSD of the light curve given the fitted CARMA model, where we opt to use 10,000 samples. Similar to our structure function analysis, we use the median value of the CARMA parameters and PSD as the best-fit value, and the 16$^{\rm th}$ and 84$^{\rm th}$ percentiles of the samples to obtain the uncertainties in the values.

In a similar manner to \cite{Simm_etal_2016}, we define a median noise level, 2 $\times$ median($\Delta t$) $\times$ median($\sigma_y^2$), for each PSD to define where the PSD is credible. In this expression, $\Delta t$ is a list of time lags in a given time-series, and $\sigma_y$ is the measurement uncertainty in the light curve fluxes.

\subsection{Sampling Methods}\label{app:Sampling}

\begin{figure}
    \centering
    \includegraphics[width=.45\textwidth]{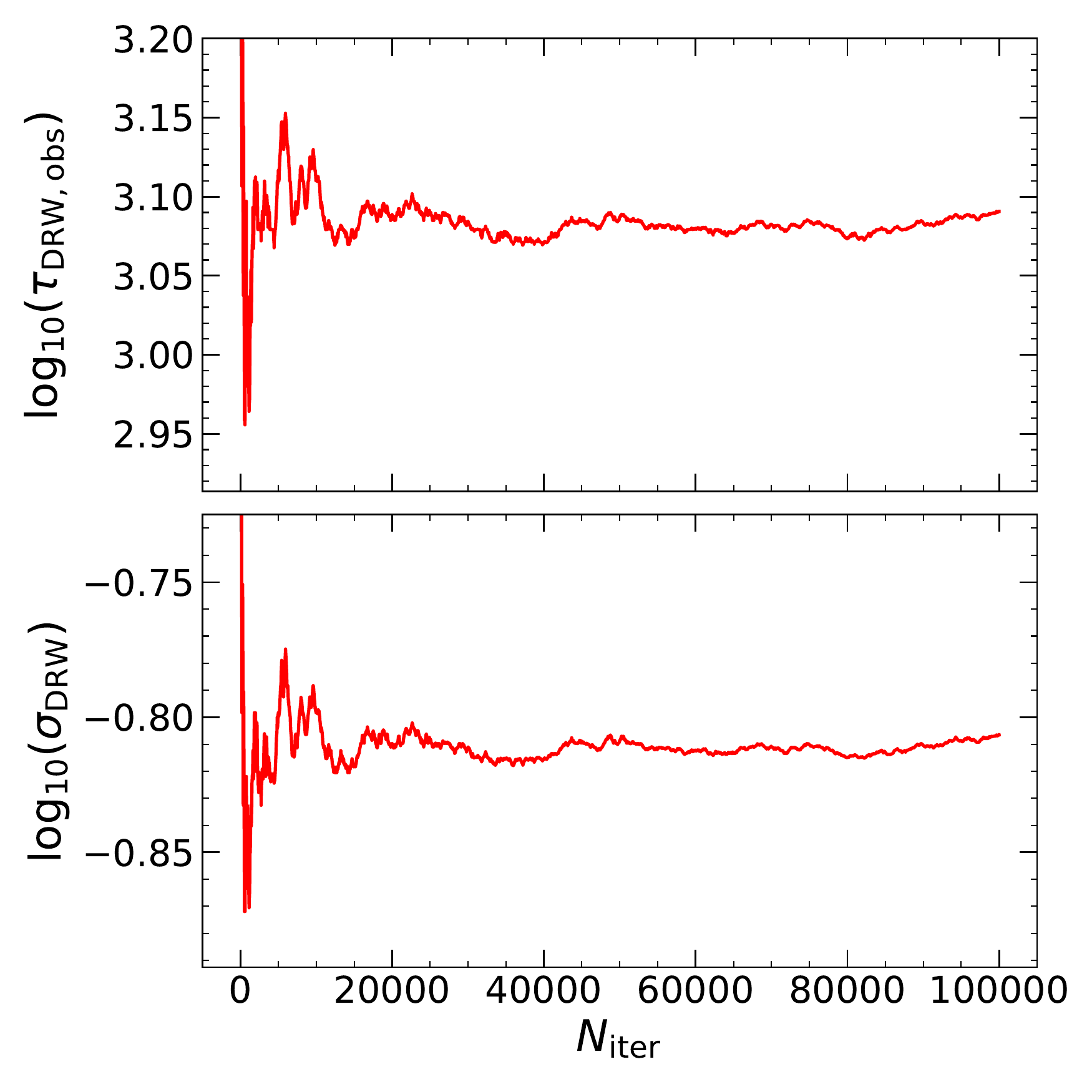}
    \caption{Convergence of fitted DRW parameters for an example quasar light curve, using a CARMA(1,0) model in {\tt CARMA\_pack}, as a function of the number of samples generated by the MCMC sampler. In this study, we opt to use $N_{\rm samp} = 60,000$, well within the range where both of these parameters cease to vary significantly.}
    \label{fig:ParamConvergence}
\end{figure}

\begin{figure}
    \centering
    \includegraphics[width=.45\textwidth]{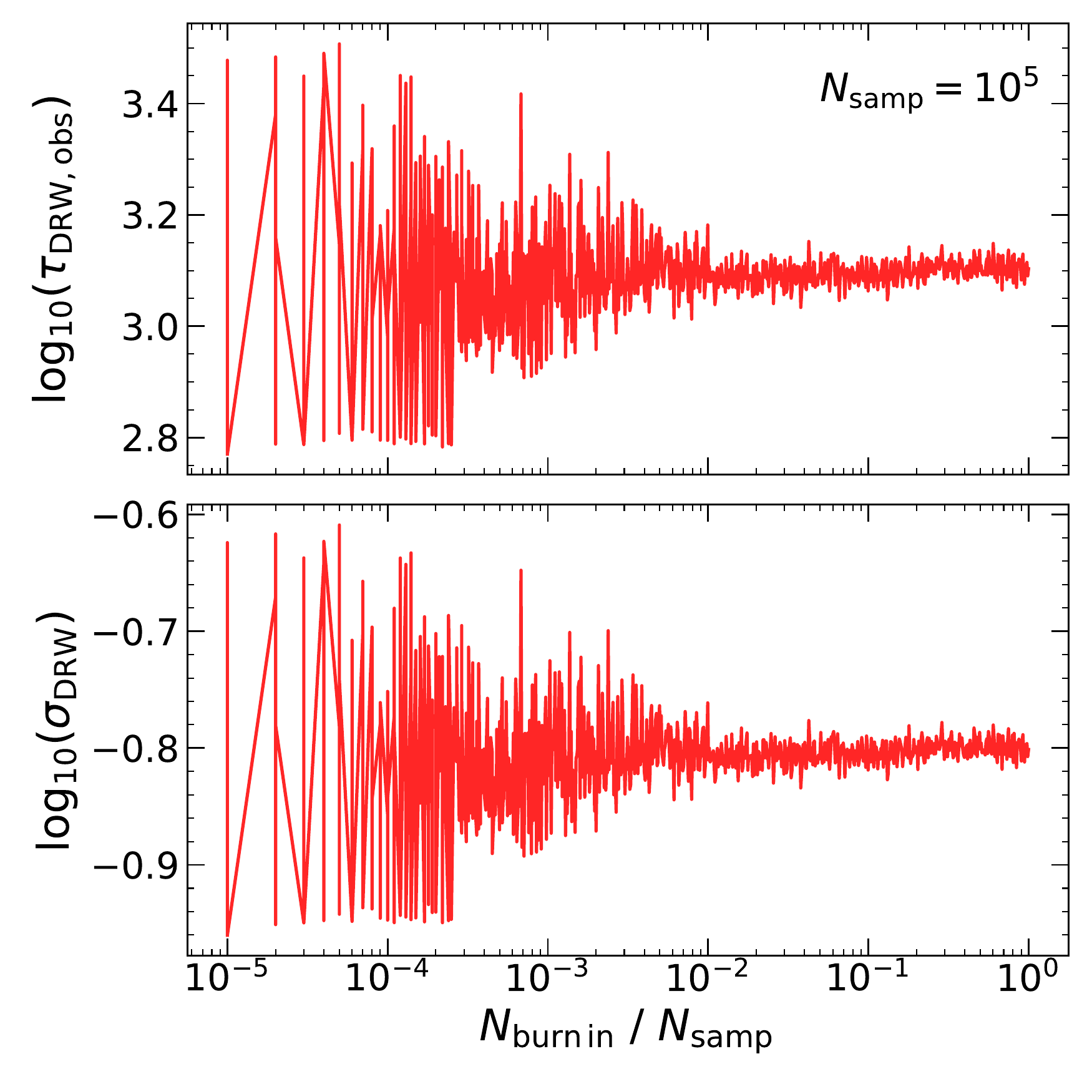}
    \caption{ DRW parameters recovered from a DRW fit to an example quasar light curve, using a CARMA(1,0) model in {\tt CARMA\_pack}, as the number of burn-in samples for the MCMC sampler increases. We choose $N_{\rm samp} = 10^5$, as Fig.~\ref{fig:ParamConvergence} showed that for this number of iterations, both parameters have already converged. We opt to use the default number of burn-in samples that {\tt CARMA\_pack} chooses ($0.5 N_{\rm samp}$) for which the output parameters have already converged. }
    \label{fig:BurnIn}
\end{figure}

One significant step in generating DRW and CARMA parameters for each light curve is the generation of samples from the posterior probability distribution through the use of MCMC sampling. In {\tt Celerite}, this is done using the popular python-based MCMC sampler {\tt emcee}, while the sampling in {\tt CARMA\_pack} is done through a custom, C++ MCMC sampler. One important parameter of sampling is the number of burn-in samples and actual samples to use for a given dataset. The burn-in samples for an MCMC sampler help to initialize the sampler to the data and allow it to converge properly. The number of actual samples for an MCMC sampler affects how well the posterior probability distribution for a parameter is sampled. For {\tt Celerite} DRW fits, we opt to use 500 burn-in samples and 2,000 actual samples, which we found to be the optimal values through trial and error. For the {\tt CARMA\_pack} fits, we adjust the number of burn-in samples relative to the total samples as well as the number of total samples to see where the results from the sample would converge and have low fluctuations. In Fig.~\ref{fig:ParamConvergence} we show the evolution of the two DRW parameters over iterations of the sampler when fitting one of our quasar light curves to a CARMA(1,0) model in {\tt CARMA\_pack}. We can see that there are large fluctuations in the sampled value in the early iterations, but the value converges to a set value after $\sim50,000$ iterations of the sampler. In Fig.~\ref{fig:BurnIn} we show the evolution of the sampled parameter values for these DRW parameters, for the same quasar, as the number of burn-in samples increases for a fixed total number of samples of 100,000 (which we have seen has a converged parameter value). This shows that the uncertainty of the value produced with a relatively low number of burn-in samples is high, but decreases to a nearly constant value at $\sim2,000$ burn-in samples (.02 $N_{\rm burn \, in} / N_{\rm samp}$). The default value for the number of burn-in samples is half the total number of samples, which adequately allows for the initialization of the sampler. Therefore, we opt to use 60,000 samples and 30,000 burn-in samples for each {\tt CARMA\_pack} fit.

\subsection{Fitting simulated DRW light curves}\label{app:mock_lc}


\begin{figure}
    \centering
    \includegraphics[width=.4\textwidth]{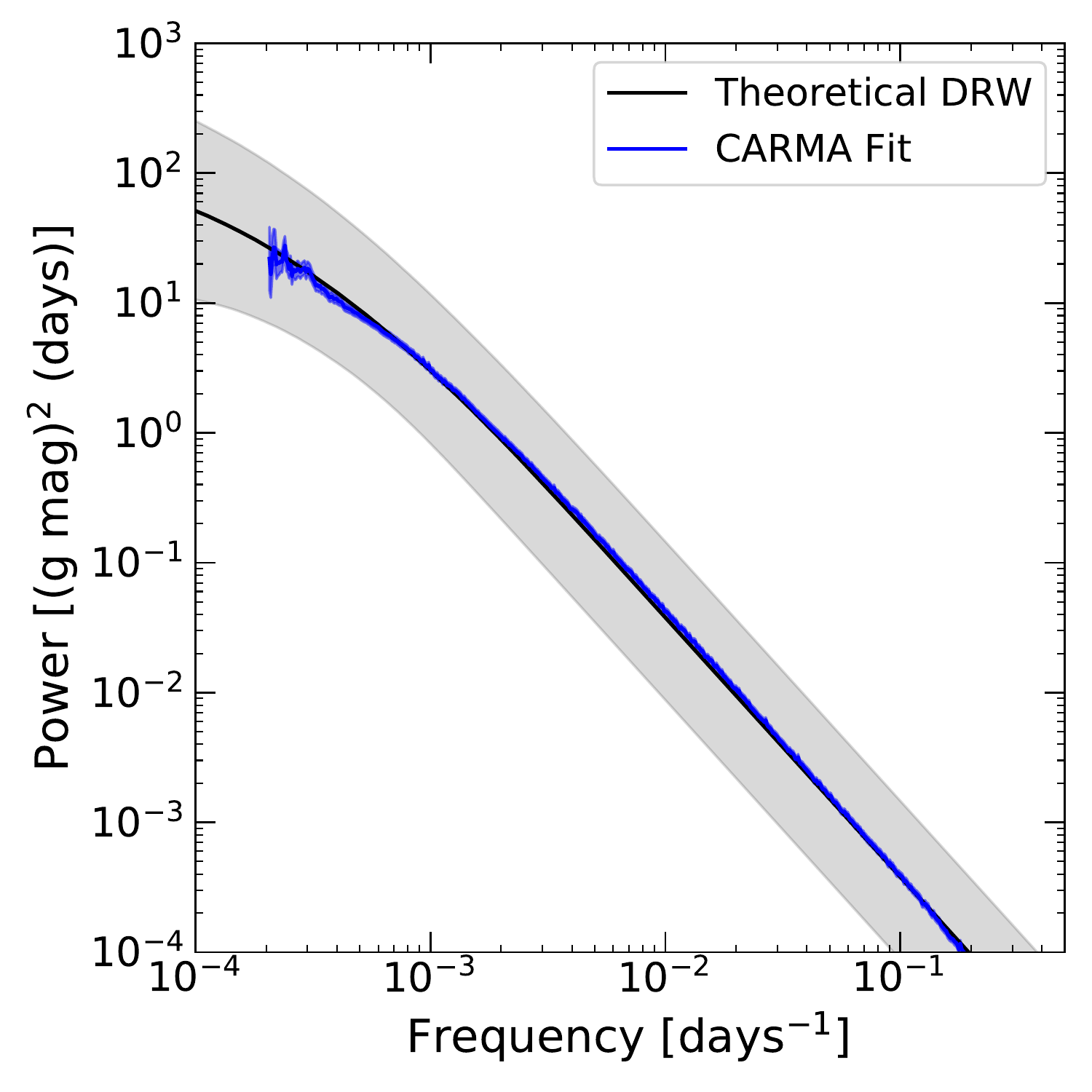}
    \caption{The ensemble PSD from mock DRW light curves generated using the best-fit DRW parameters from the real quasar light curves in \S\ref{sec:drw}. The expected DRW ensemble (median) PSD is shown in the black solid line with 1$\sigma$ uncertainties highlighted in gray. The blue line shows the ensemble (median) PSD from generalized CARMA fits with {\tt CARMA\_pack}. The generalized CARMA fits correctly recover the DRW PSD, with no evidence for slope steepening at the highest frequencies sampled here.  }
    \label{fig:CompareSim}
\end{figure}


Here we test if simulated DRW light curves with the same sampling and S/N as our real data would produce a steep high-frequency slope in the CARMA PSD. First, we generate mock $g$-band DRW light curves for all quasars in our sample using the best-fit $\tau_{\rm DRW, obs}$ and $\sigma_{\rm DRW}$ in \S\ref{sec:drw}. These mock light curves are sampled at the same times and with the same S/N as the real light curves in our sample. 


Next, we use {\tt CARMA\_pack} to fit these mock DRW light curves with a generalized CARMA model following the same procedures described in Appendix \ref{app:psd}. We then extract a PSD from each mock light curve from the best-fit CARMA model, and construct an ensemble PSD. The results are shown in Fig.~\ref{fig:CompareSim}, where we compare the PSDs from the expected DRW model and recovered by {\tt CARMA\_pack}. We find that {\tt CARMA\_pack} successfully recovers a DRW PSD for these simulated light curves, as expected. This test confirms that the steep high-frequency-end PSD slope seen in real data is not due to effects of light curve cadence and S/N or the use of a more flexible CARMA model to fit the light curves. 


We also use these simulated DRW light curves to investigate different choices of the best-fit parameters in {\tt Celerite} or {\tt CARMA\_pack}. In this work, we opt to use the median of the posterior distribution of samples as the fiducial best-fit parameters for all DRW and general CARMA model fits. Other works may use different choices for their best-fit parameters (such as the maximum-a-posteriori (MAP) \citep{MacLeod_etal_2010} or the expectation value of the marginalized posterior \citep{Suberlak_etal_2021}). Here, we discuss the differences in these choices of the best-fit parameters.

When modeling our quasar light curves with {\tt Celerite} or {\tt CARMA\_pack}, we are given a number of samples for each parameter, output by a certain MCMC algorithm. The posterior probability distribution is simply the normalized distribution of the output parameters themselves. Using the median of the posterior is less susceptible to large fluctuations in the probability due to insufficient sampling of the distribution. The MAP, however, can prove to be unreliable, as it can be easily influenced by these fluctuations. The marginalized posterior utilizes the joint-posterior distribution of multiple parameters, giving a more robust look into the relationships between parameters, and taking that into account to choose the best possible value. The expectation value of this distribution (as opposed to the MAP or median) can aid if the posterior distribution has multiple peaks.

We compare different choices, including: median posterior, MAP, and expectation value. Both the MAP and the expectation value of a parameter's distribution are obtained by using the marginalized distribution of each parameter. This is done through the use of the likelihoods output from the {\tt Celerite} fitting, for each quasar. One of the functions implemented in {\tt Celerite} allows one to obtain a likelihood for a given set of parameters and data, given the model fit to a certain set of data. Therefore, for each sample from a given quasar light curve fit, we can construct a grid in parameter space, performing this likelihood calculation for an arbitrary number of points to obtain an n-dimensional posterior distribution, where n is the number of parameters. In this case, {\tt Celerite} fits for both DRW parameters and a noise term, making this posterior three-dimensional. We can then marginalize over this distribution for each of the parameters, and obtain a best-fit parameter for each light curve. 


\begin{figure*}
    \centering
    \includegraphics[width=.9\textwidth]{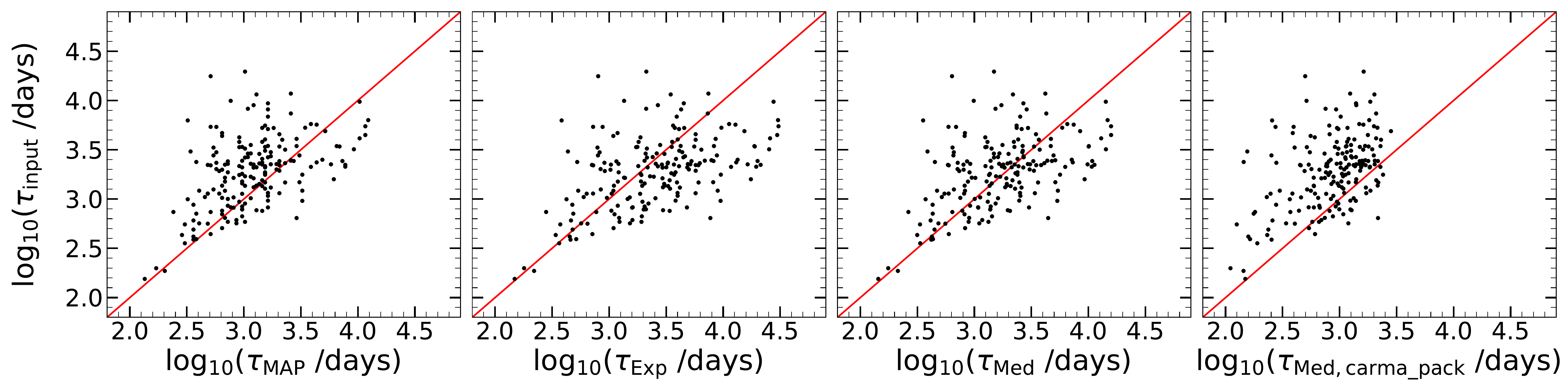}
    \includegraphics[width=.9\textwidth]{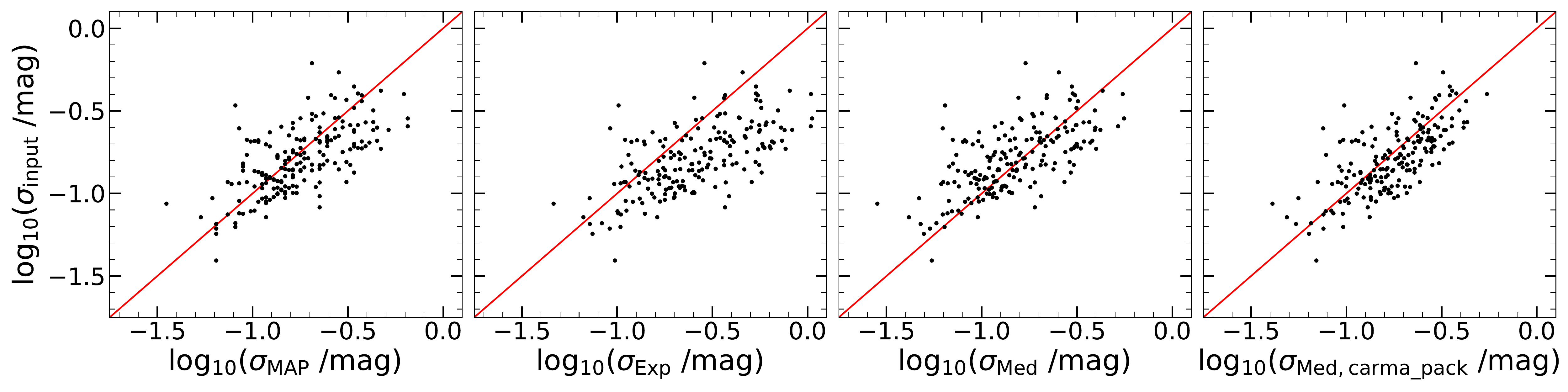}
    \caption{Comparison of the recovered and input DRW parameters ($\tau_{\rm DRW, obs}$ , $\sigma_{\rm DRW}$) from our test using simulated DRW light curves. The best-fit value for the recovered parameter was obtained via three different methods: (1) the MAP from {\tt Celerite} fitting samples, (2) the expectation value of the marginalized posterior using {\tt Celerite} fitting samples, (3) the median value of the {\tt Celerite} fitting samples, and (4) the median value of the {\tt CARMA\_pack} DRW fit samples. For each panel, the unity relation is shown as a red line. }
    \label{fig:CompareBestfit}
\end{figure*}

\begin{figure*}
    \centering
    \includegraphics[width=.6\textwidth]{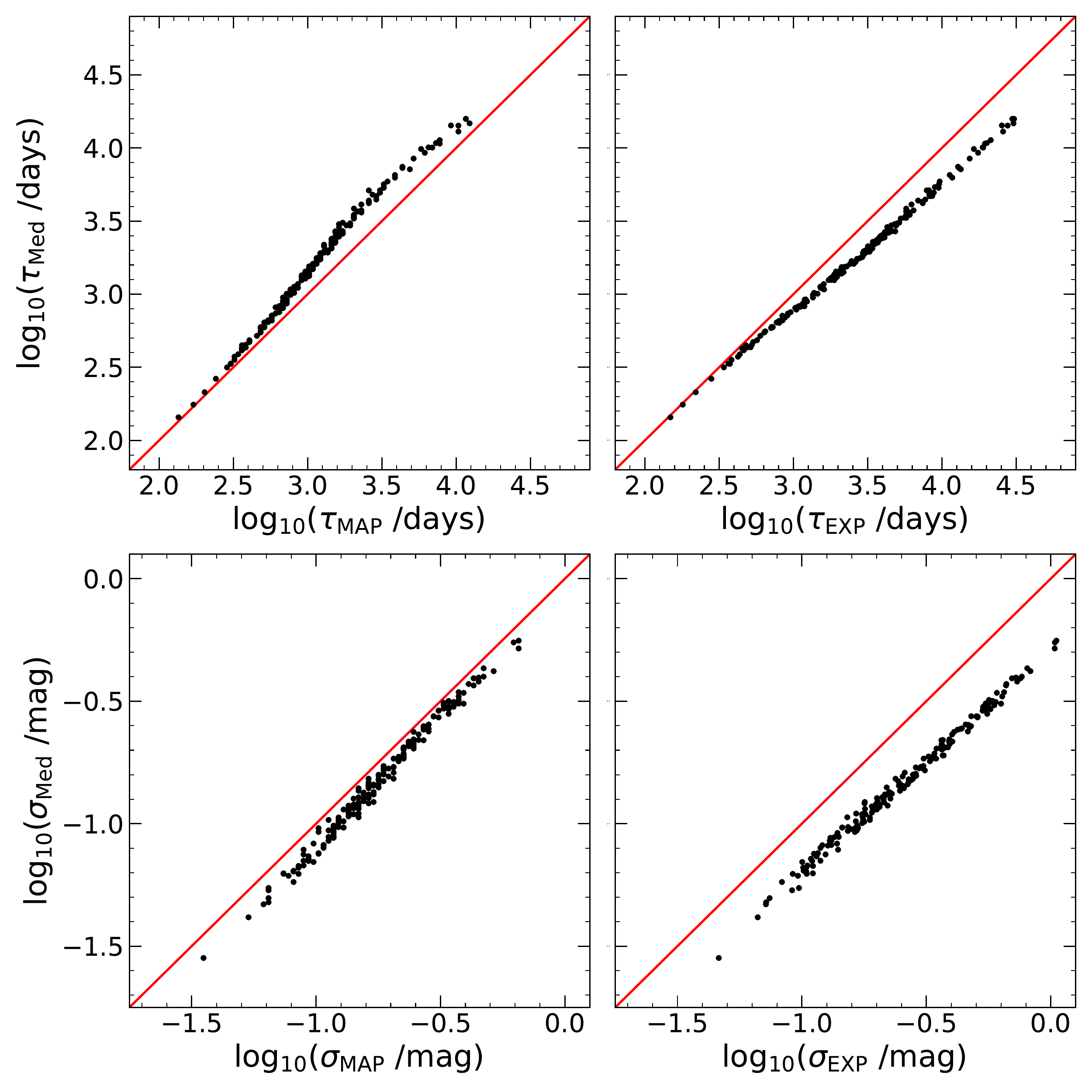}
    \caption{Comparison of different choices of the best-fit DRW parameters in {\tt Celerite}, obtained from our simulated DRW light curves. These $\tau_{\rm DRW, obs}$ and $\sigma_{\rm DRW}$ values are determined using the same sets of posterior samples. The method used to obtain these MAP values and expectation values uses the marginalized posterior distribution of the samples. The red line in each panel indicates the unity relation.}
    \label{fig:CompareBestfit_celerite}
\end{figure*}

We compare these different choices of best-fit parameters in Fig.~\ref{fig:CompareBestfit}. We obtained these values from fitting our simulated DRW light curves with a DRW model with {\tt Celerite}, as well as {\tt CARMA\_pack} (in the latter case, a DRW or CARMA(1,0) model is enforced). We find that all these choices perform similarly for both DRW parameters with a similar amount of scatter, when compared with the input DRW parameters used to construct the simulated light curves. Overall, $\sigma_{\rm DRW}$ is better recovered than $\tau_{\rm DRW, obs}$. For long input $\tau_{\rm DRW}$, the recovered $\tau_{\rm DRW}$ is generally biased low due to an insufficient baseline of the light curve \citep[e.g.,][]{Kozlowski_2017}. Interestingly, the median posterior from the {\tt Celerite} fit produces the least overall bias in $\tau_{\rm DRW}$ for our sample, justifying our choice of this particular definition of best-fit DRW parameters in this work. 


Fig.~\ref{fig:CompareBestfit_celerite} shows the comparison of the three different choices of the best-fit DRW parameters, using {\tt Celerite} for the same simulated DRW light curves described above. While there are correlations among these different choices, there are also systematic offsets among them. For this study, we have chosen the median posterior as our fiducial best-fit parameters, given its performance in recovering the input DRW parameters as demonstrated in Fig.~\ref{fig:CompareBestfit}.



\section*{Affiliations}
{\small \textit{
$^{1}$Department of Astronomy, University of Illinois at Urbana-Champaign, Urbana, IL 61801, USA\\
$^{2}$National Center for Supercomputing Applications, University of Illinois at Urbana-Champaign, Urbana, IL 61801, USA\\
$^{3}$Center for AstroPhysical Surveys, National Center for Supercomputing Applications, University of Illinois at Urbana-Champaign, Urbana, IL 61801, USA\\
$^{4}$Harvard-Smithsonian Center for Astrophysics,  60 Garden Street, Cambridge, MA 02138, USA\\
$^{5}$Instituto de F\'{i}sica Te\'orica, Universidade Estadual Paulista, S\~ao Paulo, Brazil\\
$^{6}$Laborat\'orio Interinstitucional de e-Astronomia - LIneA, Rua Gal. Jos\'e Cristino 77, Rio de Janeiro, RJ - 20921-400, Brazil\\
$^{7}$Fermi National Accelerator Laboratory, P. O. Box 500, Batavia, IL 60510, USA\\
$^{8}$Institute of Cosmology and Gravitation, University of Portsmouth, Dennis Sciama Building, Burnaby Road, Portsmouth, PO1 3FX, UK\\
$^{9}$Canada-France-Hawaii Telescope Corporation, 65-1238 Mamaloahoa Hwy, Kamuela, HI 96743, USA\\
$^{10}$Sorbonne Universit\'e, CNRS, UMR 7095, Institut d'Astrophysique de Paris, 98 bis boulevard Arago, F-75014 Paris, France\\
$^{11}$Faculty of Physics, Ludwig-Maximilians-Universit\"at, Scheinerstr. 1, 81679 Munich, Germany\\
$^{12}$Department of Physics \& Astronomy, University College London, Gower Street, London, WC1E 6BT, UK\\
$^{13}$Kavli Institute for Particle Astrophysics \& Cosmology, P. O. Box 2450, Stanford University, Stanford, CA 94305, USA\\
$^{14}$SLAC National Accelerator Laboratory, Menlo Park, CA 94025, USA\\
$^{15}$Institut de F\'{\i}sica d'Altes Energies (IFAE), The Barcelona Institute of Science and Technology, Campus UAB, 08193 Bellaterra (Barcelona) Spain\\
$^{16}$Port d'Informaci\'{o} Cient\'{i}fica (PIC), Campus UAB, C. Albareda s/n, 08193 Bellaterra (Cerdanyola del Vall\`{e}s), Spain\\
$^{17}$Observat\'orio Nacional, Rua Gal. Jos\'e Cristino 77, Rio de Janeiro, RJ - 20921-400, Brazil\\
$^{18}$Department of Physics, University of Michigan, Ann Arbor, MI 48109, USA\\
$^{19}$Hamburger Sternwarte, Universit\"{a}t Hamburg, Gojenbergsweg 112, 21029 Hamburg, Germany\\
$^{20}$Centro de Investigaciones Energ\'eticas, Medioambientales y Tecnol\'ogicas (CIEMAT), Madrid, Spain\\
$^{21}$Department of Physics, IIT Hyderabad, Kandi, Telangana 502285, India\\
$^{22}$Institute of Theoretical Astrophysics, University of Oslo. P.O. Box 1029 Blindern, NO-0315 Oslo, Norway\\
$^{23}$Instituto de F\'isica Te\'orica UAM/CSIC, Universidad Aut\'onoma de Madrid, 28049 Madrid, Spain\\
$^{24}$Institut d'Estudis Espacials de Catalunya (IEEC), 08034 Barcelona, Spain\\
$^{25}$Institute of Space Sciences (ICE, CSIC),  Campus UAB, Carrer de Can Magrans, s/n,  08193 Barcelona, Spain\\
$^{26}$Faculty of Physics, Ludwig-Maximilians-Universit\"at, Scheinerstr. 1, 81679 Munich, Germany\\
$^{27}$School of Mathematics and Physics, University of Queensland,  Brisbane, QLD 4072, Australia\\
$^{28}$Santa Cruz Institute for Particle Physics, Santa Cruz, CA 95064, USA\\
$^{29}$Center for Cosmology and Astro-Particle Physics, The Ohio State University, Columbus, OH 43210, USA\\
$^{30}$Department of Physics, The Ohio State University, Columbus, OH 43210, USA\\
$^{31}$Australian Astronomical Optics, Macquarie University, North Ryde, NSW 2113, Australia\\
$^{32}$Lowell Observatory, 1400 Mars Hill Rd, Flagstaff, AZ 86001, USA\\
$^{33}$Centre for Gravitational Astrophysics, College of Science, The Australian National University, ACT 2601, Australia\\
$^{34}$The Research School of Astronomy and Astrophysics, Australian National University, ACT 2601, Australia\\
$^{35}$Instituci\'o Catalana de Recerca i Estudis Avan\c{c}ats, E-08010 Barcelona, Spain\\
$^{36}$Institut de F\'{\i}sica d'Altes Energies (IFAE), The Barcelona Institute of Science and Technology, Campus UAB, 08193 Bellaterra (Barcelona) Spain\\
$^{37}$Physics Department, University of Wisconsin-Madison, Madison, WI, USA\\
$^{38}$Institute of Astronomy, University of Cambridge, Madingley Road, Cambridge CB3 0HA, UK\\
$^{39}$Department of Astrophysical Sciences, Princeton University, Peyton Hall, Princeton, NJ 08544, USA\\
$^{40}$School of Physics and Astronomy, University of Southampton,  Southampton, SO17 1BJ, UK\\
$^{41}$Computer Science and Mathematics Division, Oak Ridge National Laboratory, Oak Ridge, TN 37831\\
$^{42}$Waldorf High School of Massachusetts Bay, Belmont, MA 02478, USA
}}

\end{appendix}


\bsp	
\label{lastpage}
\end{document}